\documentclass[sigconf]{acmart}
\usepackage{makecell}
\usepackage{paralist}
\usepackage{dblfloatfix}
\usepackage{color, colortbl}

    \usepackage{graphicx}
    \usepackage[normalem]{ulem}
    \useunder{\uline}{\ul}{}

\usepackage{tabularray}
\usepackage{caption}
\usepackage{subcaption}
\usepackage{csquotes}

\AtBeginDocument{%
  \providecommand\BibTeX{{%
    \normalfont B\kern-0.5em{\scshape i\kern-0.25em b}\kern-0.8em\TeX}}}

\copyrightyear{2023}
\acmYear{2023}
\setcopyright{rightsretained}
\acmConference[ASSETS '23]{The 25th International ACM SIGACCESS Conference on Computers and Accessibility}{October 22--25, 2023}{New York, NY, USA}
\acmBooktitle{The 25th International ACM SIGACCESS Conference on Computers and Accessibility (ASSETS '23), October 22--25, 2023, New York, NY, USA}
\acmDOI{10.1145/3597638.3608386}
\acmISBN{979-8-4007-0220-4/23/10}
\begin{document}

\title["The Guide Has Your Back"]{"The Guide Has Your Back": Exploring How Sighted Guides Can Enhance Accessibility in Social Virtual Reality for Blind and Low Vision People}

\author{Jazmin Collins}
\authornote{Jazmin Collins and Crescentia Jung are co-authors and contributed equally to this research.}
\email{jc2884@cornell.edu}
\affiliation{%
  \institution{Cornell University}
  \streetaddress{2 W Loop Rd}
  \city{New York}
  \state{NY}
  \country{USA}
  \postcode{10044}
}
\author{Crescentia Jung}
\authornotemark[1]
\email{cj382@cornell.edu}
\affiliation{%
  \institution{Cornell University}
  \streetaddress{2 W Loop Rd}
  \city{New York}
  \state{NY}
  \country{USA}
  \postcode{10044}
}
\author{Yeonju Jang}
\email{yj376@cornell.edu}
\affiliation{%
  \institution{Cornell University}
  \streetaddress{}
  \city{Ithaca}
  \state{NY}
  \country{USA}
  \postcode{14850}
}
\author{Danielle Montour}
\email{danielle.mountour1@gmail.com}
\affiliation{%
  \institution{Cornell University}
  \streetaddress{2 W Loop Rd}
  \city{New York}
  \state{NY}
  \country{USA}
  \postcode{10044}
}
\author{Andrea Stevenson Won}
\email{asw248@cornell.edu}
\affiliation{%
  \institution{Cornell University}
  \streetaddress{}
  \city{Ithaca}
  \state{NY}
  \country{USA}
  \postcode{14850}
}
\author{Shiri Azenkot}
\email{shiri.azenkot@cornell.edu}
\affiliation{%
  \institution{Cornell Tech}
  \streetaddress{2 W Loop Rd}
  \city{New York}
  \state{NY}
  \country{USA}
  \postcode{10044}
}

\renewcommand{\shortauthors}{Collins and Jung, et al.}

\begin{abstract}
As social VR applications grow in popularity, blind and low vision users encounter continued accessibility barriers. Yet social VR, which enables multiple people to engage in the same virtual space, presents a unique opportunity to allow other people to support a user’s access needs. To explore this opportunity, we designed a framework based on physical sighted guidance that enables a guide to support a blind or low vision user with navigation and visual interpretation. A user can virtually hold on to their guide and move with them, while the guide can describe the environment. We studied the use of our framework with 16 blind and low vision participants and found that they had a wide range of preferences. For example, we found that participants wanted to use their guide to support social interactions and establish a human connection with a human-appearing guide. We also highlight opportunities for novel guidance abilities in VR, such as dynamically altering an inaccessible environment. Through this work, we open a novel design space for a versatile approach for making VR fully accessible.
\end{abstract}
\nopagebreak





\maketitle

\section{Introduction} \label{Introduction}

As consumer virtual reality headsets grow in popularity and drop in price \cite{what-is-the-metaverse}, virtual reality (VR) is increasingly used for social interaction \cite{mcveigh-schultz-mapping-social-vr-design-ecology}. The social VR application VRChat is consistently one of the most played VR applications on Steam, several hundred spots above well-known single-player experiences like the game Beat Saber \cite{steam-charts-top-played-games}. However, despite this growing interest in VR and social VR in particular, mainstream designs are generally biased toward visual content, thus excluding blind and low vision (BLV) users. Unless effective measures are developed to make VR accessible, millions of BLV people will continue to be marginalized or altogether absent in VR spaces as they become more prevalent.

Researchers have enhanced virtual environments with added audio and haptic effects to make objects and navigation more accessible. For example, Zhao et al. \cite{zhao-Canetroller} and Siu et al. \cite{siu-VR-without-vision} developed a custom controller that simulated navigation with a white cane; it provided force feedback and sound effects when it hit virtual objects. Other projects leveraged audio and haptics on VR devices themselves. Virtual Showdown \cite{wedoff-virtualshowdown} and AudioDoom \cite{audiodoom} provide audio cues to signify the location and identity of objects in the environment, as well as provide feedback for user actions. While such projects improve VR accessibility to some extent, they were mostly developed for simple, single-user experiences with specific goals. With the rising prominence of social VR, researchers must consider more versatile accessibility measures for complex, multiuser VR spaces like VRChat and Horizon Worlds. To the best of our knowledge, prior work has not considered such environments. 

In the physical world, BLV people often use guides or visual interpreters to manage accessibility needs in unfamiliar environments. Yet these experiences cannot currently be applied in VR. For example, in the physical world, a BLV person uses a \textit{sighted guide} \cite{sightedguide} by touching the back of a sighted person’s elbow and walking alongside and slightly behind them. However, even though social VR allows multiple users to engage in the same virtual space, there is no way to obtain assistance in today’s mainstream virtual experiences. In addition, VR provides affordances (e.g., flying, invisibility) that are unavailable in the physical world, creating unique opportunities for virtual guidance.

In this paper, we explore how a physical guide’s functions can be implemented and extended in VR to support overall accessibility for BLV users. Specifically, we address two research questions:

\begin{itemize}
\item RQ1: What accessibility needs can a guide address for BLV people in VR?
\item RQ2: What design approaches for a guide address these needs?
\end{itemize}

To answer these questions, we conducted a study with 16 BLV participants who used a VR prototype that emulated sighted guidance in the physical world. We designed the prototype with minimal guidance scaffolding to give participants agency and flexibility while interacting with the guide. We developed three simulated park environments, each with several interactable objects and obstacles as well as agent-avatars. One researcher acted as a sighted guide and entered the parks with participants as they explored. Our goal was to observe their use patterns and challenges with the guide. We concluded each session with an interview to elicit participants’ reflections on their experience with the guide and ways in which it could be improved.

We found that guides can effectively address a variety of BLV people’s needs, including supporting mobility and orientation, describing visuals, and providing companionship. Participants had differing opinions about a guide’s ideal level of initiative, functions, forms, and visibility. For instance, some participants wanted to establish a human connection with a human-appearing guide and extend their guide’s support to social interactions. Others wanted to be more independent of the guide. Some even wanted invisible guides so no one could tell they were using one. Finally, we discuss opportunities for exploring guidance frameworks, such as crowdsourced or friend-sourced guides. We also outline opportunities for novel guidance abilities in VR, such as dynamically altering an inaccessible environment or temporarily controlling participants’ actions in VR.

To summarize, we contribute the first exploration of using a sighted guide to enhance accessibility in VR. We provide an in-depth analysis of BLV people’s experience with a virtual guide and outline opportunities for future exploration of virtual guidance.

\section{Related Work} \label{Related Work}
\subsection{Enhancing Accessibility of VR} \label{Enhancing Accessibility of VR}

To enhance the accessibility of VR, researchers have developed ways to convey information about the environment layout and objects using audio and haptics. Most prior work has developed specialized tools and environments for BLV people, while much less has focused on making mainstream VR applications accessible. 

Many researchers have explored accessible navigation in VR \cite{andrade-echo-house, nair-navstick, hao-detect-and-approach-navigation, guerreiro2023design}. For instance, Andrade et al. created EchoHouse, a virtual environment based on echolocation where objects emitted real-time audio feedback to provide a mental map of the space. Other researchers have used audio and haptics to simulate the touch-based feedback of a white cane for virtual obstacle detection \cite{zhao-Canetroller, siu-VR-without-vision, tzovaras-mr-white-cane, kim-vivr-cane, zhang-exploring-ve-mixed-reality-cane}. For example, Zhao et al. \cite{zhao-Canetroller} created the Canetroller, a handheld device that uses audio and haptics to emulate the sensations of a white cane running into objects or dragging across different surfaces. Others have used haptics to simulate force feedback from gripping objects \cite{nikolakis-cybergrasp, kornbrot-roughness, sinclair-capstan-crunch}, such as Sinclair et al.’s Capstan Crunch \cite{sinclair-capstan-crunch}, or as an experimental method of requesting object information via haptic gloves \cite{penuela-hands-on}. However, these projects require custom hardware, which can make their solutions difficult to adopt for widespread use.

In addition, researchers have developed various specialized VR games for BLV people that use a combination of audio and haptic feedback to make game content perceivable \cite{gluck-racing-in-the-dark, gluck-the-enclosing-dark, wedoff-virtualshowdown, audiodoom, morelli_vi-tennis_2010, blindswordsman}. For example, Wedoff et al.’s Virtual Showdown \cite{wedoff-virtualshowdown} is an accessible VR game that uses audio, verbal, and haptic feedback to locate and hit a ball against an opponent.

	Only a few researchers have examined mainstream VR, creating methods for VR developers to integrate accessibility into their applications \cite{zhao2019seeingvr, li-soundvizvr}. One prominent example is Zhao et al.’s SeeingVR, a Unity toolkit containing 14 accessibility tools that enhance virtual environments for people with low vision. Each of these tools can be integrated with VR applications during or after development, transforming mainstream applications into accessible experiences \cite{zhao2019seeingvr}. Meta has also released Meta Haptics Studio \cite{meta-haptics-studio}, which allows VR developers to upload sound effects files and create customized haptic patterns.

However, all of the above work focuses on single-user VR experiences. In fact, most efforts have contributed to games, where the environment is designed for a single purpose. In contrast, in a multiuser space like Mozilla Hubs or VRChat, other people frequently modify the environment’s content for various purposes. While prior work serves as a baseline for accessible VR experiences, they are not equipped to handle the complexities of multiuser spaces. Thus, our work explores a more flexible and powerful approach with a sighted guide, leveraging the fact that social VR platforms can accommodate multiple people.

\subsection{Enhancing the Accessibility of Social Virtual Worlds} \label{Enhancing the Accessibility of Social Virtual Worlds}

Researchers have conducted work to improve accessibility in social virtual worlds before the advent of immersive, consumer VR \cite{kaltenhauser-ghosts-from-past-secondlife, white-towards-accessible-secondlife, depascale-haptics-to-secondlife, huang-3d-haptic-labyrinth, nair2022uncovering, gonccalves2023my}. Virtual worlds created through computer and browser-based experiences have become long-standing social hubs for millions of users. Notable examples include SecondLife, World of Warcraft, and Roblox, which have built enormous user communities \cite{greener-second-life-traffic, mmo-population-list}.

Kahlifa’s virtual world client Radegast is among one of the best-known accessibility efforts in SecondLife \cite{khalifa-radegast}. It is an open-source GitHub repository that provides non-visual interaction for BLV users of SecondLife. Radegast translates the 3D-graphical interface of SecondLife into a text-based game viewer, making the game compatible with screen readers for BLV users.

Accessibility has also been added to virtual worlds from development \cite{trewin-powerup, westin-terraformers}. Trewin et al. worked on the development team of a game called PowerUp, designed to be an accessible multiplayer experience for a range of physical and cognitive disabilities. The game allowed for multiple forms of user input–voice input, keyboard-only input, mouse-only input–and provided visual alterations like font customization for low vision players \cite{trewin-powerup}.

Motivated by its increased popularity, researchers have recently begun to address the accessibility of immersive, social VR for BLV users. Zhang et al. explored avatar diversity and self-presentation of people with disabilities, emphasizing the importance of disclosing one’s disability through avatars and customization \cite{zhang-part-of-me-avatar-diversity}. Ji et al.’s VRBubble was an immersive VR system with three types of audio feedback (earcons, verbal feedback, and physical sound earcons) that provided proximity information about other avatars for BLV users \cite{ji-vrbubble}. Wieland et al. conducted interviews on how BLV people use social cues in the physical world to inform how such cues could be used in social VR \cite{wieland-nvc-joint-attention-guidelines}. Gonçalves et al. explored audio feedback and proposed navigation methods like automatic walking and co-movement to facilitate group interaction \cite{gonccalves2023inclusive}. 

While efforts in social virtual worlds show the breadth of possibilities for accessible development in VR, much work remains. Text-based interfaces like Radegast \cite{khalifa-radegast}, for example, are not easy to translate into immersive VR, where 2D graphical interfaces that support text are replaced by 3D graphical elements. Such elements are typically inaccessible to text-based assistive technology like screen readers unless explicitly designed to be so \cite{guntalilib-screenreaders-vr}. Moreover, social VR presents unique challenges for BLV users, as users physically navigate in virtual environments and interact with multiple others in social contexts. Our work takes inspiration from these ventures in multiuser spaces, building on the ideals of real-time transfer of information and socially-aware accessibility methods. In order to effectively incorporate these complex functions into a single approach, we explore how a sighted guide can address existing and not-yet-identified accessibility needs.

\subsection{Visual Interpretation and Guidance in the Physical World} \label{Visual Interpretation and Guidance in the Non-Virtual World}

Visual interpretation services in the physical world are one common tool for BLV people. BLV people use their phone cameras to capture video of their surroundings; through an online application, a visual interpreter provides verbal information about one’s surroundings, helping to navigate spaces and identify objects. These services have been studied to understand users’ needs and how such services can be improved \cite{neha-be-my-eyes-aid-for-vi, avila-survey-about-be-my-eyes, nguyen-improved-quality-of-life-aira}.

One of the most well-known visual interpretation projects is VizWiz, a mobile application that connects BLV users to remote, sighted workers to answer their visual questions \cite{bigham-vizwiz}. Similarly, Aira is a paid service that provides real-time visual interpretation \cite{aira-homepage, nguyen-improved-quality-of-life-aira, lee-emerging-remote-sighted-assistance, lee-conversations-for-vision} and BeMyEyes is a free volunteer visual interpretation service \cite{be-my-eyes-homepage}. Recently, BeMyEyes introduced a new “Virtual Volunteer” feature that uses OpenAI’s GPT-4 for image-to-text interpretation. These existing services provide dynamic assistance to support the needs of BLV people in the physical world.
	
Since one strength of immersive VR lies in replicating the physical aspects of interactions, guidance in the physical world serves as a useful model. Sighted guides provide guidance for BLV people by helping them navigate spaces and offering verbal information on these spaces \cite{braille-institute-human-guide-techniques, helen-keller-human-guide}. The guide and guidee create a physical connection where the guidee grasps the guide’s elbow to feel changes in the guide’s movement. The guidee walks behind the guide as they travel, and the guide provides relevant verbal information.

All of the above services offer techniques that can inform accessible practices in social VR. However, current VR platforms lack the support to enable these techniques, with limited options for multiplayer interactions and no designs for assistance-based relationships between two users, such as allowing one player to easily guide another. As such, our work seeks to fill this gap by exploring the first of these support frameworks to connect a guide and a BLV user in VR and emphasizing opportunities for future designs. 

\section{Methods} \label{Methods}

To understand how a guide can support VR accessibility, we conducted a study with 16 BLV participants. We designed a prototype that allowed a BLV participant to explore a virtual space with a researcher acting as a sighted guide and observed the participants’ experiences. We then interviewed them to learn about their perception of the guide and ideas for ways in which the experience can be improved. Importantly, our goal was not to evaluate our prototype but rather to use it to observe and elicit participant reflections on a virtual guidance experience. 

\subsection{VR Guide Prototype} \label{VR Guide Prototype}

We developed a social VR prototype with minimal added scaffolds to enable a BLV participant to receive guidance from another user, inspired by physical sighted guidance. While many existing VR spaces allow a user to see and converse with another user, these forms of feedback are not sufficient for a BLV user to receive guidance. For example, in the physical world, a guide’s main task is to assist with navigation, which would be almost impossible without additional scaffolding in VR. Thus, our VR framework had several features to support such assistance. 

We developed our prototype in the game development engine Unity, using pre-existing scripts and assets and our own custom scripts. We deployed it on a Meta Quest 2, a consumer VR headset with two handheld controllers. 

\begin{figure*}
\includegraphics[width=450pt]{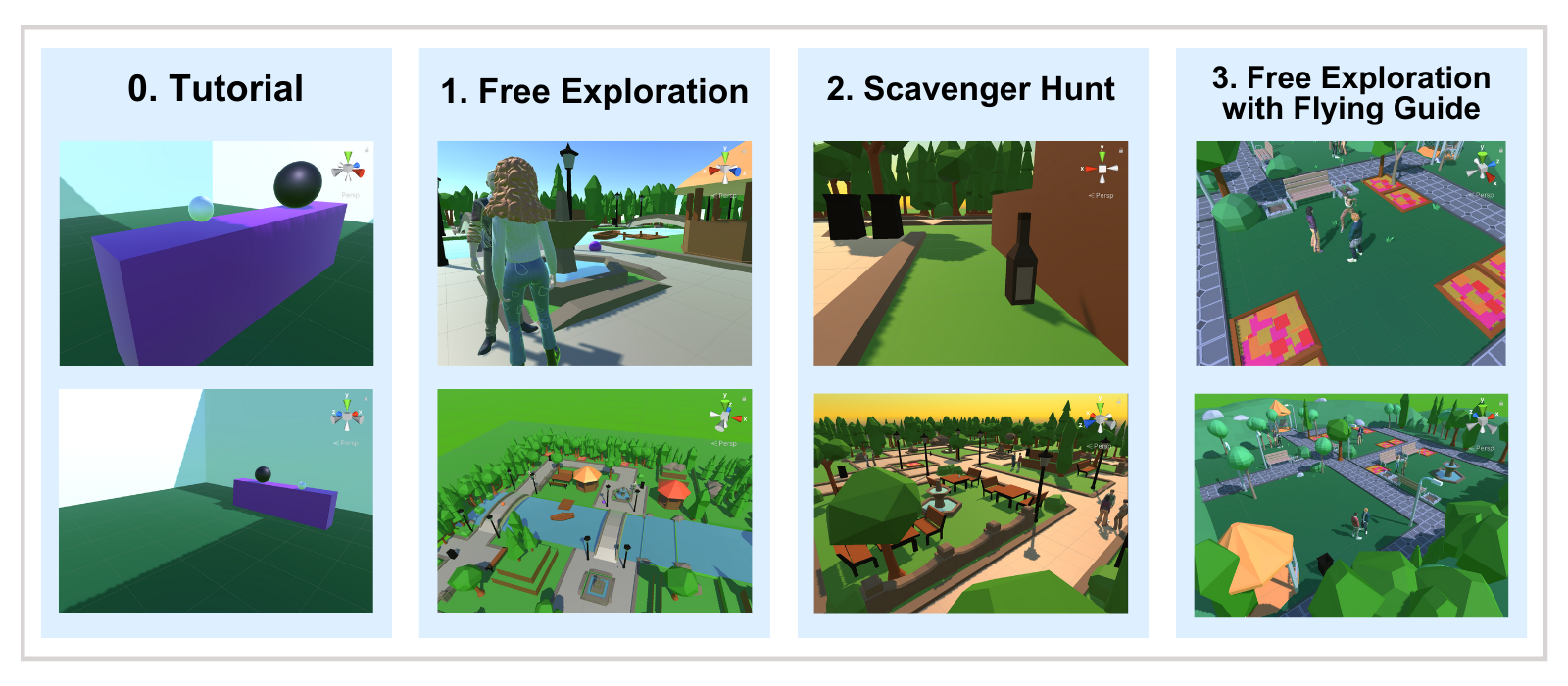}
\caption{\label{fig: Screenshots of the Tutorial and Three Park Environments from Our Prototype.} Screenshots of the Tutorial and Three Park Environments.}
\end{figure*}

\subsubsection{Designing a Representative Environment} \label{Basic Functionality for a Representative VR Environment}

Before including scaffolds to support guidance, we designed a VR environment that could simulate current social VR platforms. To determine what features the environment should include, two researchers explored existing social VR applications based on the rankings in the Oculus Quest Store and community forums \cite{xr-today-best-social-vr, ar-post-best-vr-for-socializing}. We explored three applications (RecRoom, VRChat, and Horizon Worlds) to learn the common functions and designs of current social VR experiences. Based on this exploration, we generated the following list of functions for users in the environment:
\begin{itemize}
\item Walking: participants walked by pressing the left joystick. We played the sound of footsteps.  
\item Snap turns: participants turned by flicking the right joystick in the turn direction. We played a swish sound.
\item Teleportation: participants teleported by pressing the trigger buttons on either controller and pointing to the target direction. We played an earcon.
\item Grabbing: participants grabbed objects by pressing the grip button on either controller. We played an earcon.
\end{itemize}
Finally, we integrated voice chat and a multiplayer network. We spatialized the voice from each avatar so users would hear audio coming from the other user’s virtual body. 

\subsubsection{Designing Guidance Scaffolds} \label{Scaffolds to Facilitate Guidance}

In order to simulate physical interactions between a sighted guide and a BLV user, we developed a function called \textit{Shared Movement}. During Shared Movement, the user’s avatar moves wherever the guide moves and continues until the user lets go of the guide. To initiate Shared Movement, a user must “grab” the area around a guide’s avatar by holding down the trigger while standing within 1 ft of the guide. To maintain Shared Movement, users continued to hold down the trigger. The user stops when the user releases the guide. Note that while Shared Movement is active, users cannot walk, snap turn, or perform any other movements. 

While Shared Movement was inspired by physical sighted guiding, we wanted to take advantage of transformed social interaction in VR \cite{bailenson2004transformed}. What if a sighted guide had superpowers? What if they could extend their powers to the guidee? Thus, we implemented an experimental flying function for the guide. This allowed the guide to fly off the ground and anywhere around the virtual scene, quickly gathering visual information for the user. The user could also fly with the guide by holding onto them. The guide held down the primary button on the right controller and tilted the controller to move their body. To signify this action, we played the sound of wings flapping.

\subsubsection{Additional Accessibility Enhancements} \label{Additional Accessibility Enhancements}

We completed a short formative assessment of our prototype with two of our co-authors who identify as BLV and found that it was still difficult to perform low-level tasks despite the guide. For example, they felt that basic movement functions were challenging because they lacked audio feedback. Thus, we added three audio cues for hearing the guide’s movement, the participant’s snap turns, and collisions. Additionally, we customized each of the movement-based audio cues to the surface materials of the virtual environment (e.g., wood, water, grass). Finally, we added sound effects for features in the environment (e.g., fountains and rivers).

\subsection{Behavior Guidelines for the Guide} \label{Behavior Guidelines for the Guide}

After finalizing the prototype, we established guidelines for guide’s behavior during the study. Since we wanted to uncover participant needs rather than evaluate our prototype, the guide was instructed to be fairly passive, allowing participants to control the interactions. We restricted the guide’s behavior to the following: 
\begin{itemize}
\item Asking open-ended questions to elicit instruction. For example: “How can I help?”
\item Answering user questions. If the participant asked the guide to describe the space, the guide could say, “We’re standing in a park in front of a river…” The guide would start with basic details, and only add more if requested by the participant (eg. “What else?”).
\item Asking clarifying questions to understand the participant’s goals. If the participant told the guide they would like to go to the river, the guide could ask, “Would you like me to take you there?”
\item Fulfilling participants’ requests. If the participant asked the guide to take them to the river, the guide could say, “Sure, grab on to me and we’ll go together.” After the participant engaged in Shared Movement, the guide could take them to the river.
\end{itemize}

The guide could also ask clarifying questions if they did not understand a request, but she was told to be careful not to take initiative in these exchanges. In general, the guide was told to act as though she was not familiar with the environments or study tasks and to encourage participant agency as the leader of the experience.

\begin{table*}
\centering
\begin{tblr}{
  width = \linewidth,
  colspec = {Q[100]Q[25]Q[27]Q[90]Q[148]Q[208]Q[231]Q[103]},
  hlines,
  vlines,
}
\textbf{Pseudonym} & \textbf{Age} & \textbf{Gen.} & \textbf{Vision} & \textbf{Onset}      & \textbf{Visual Acuity} & \textbf{Visual Field}            & \textbf{Mobility Aids} \\
Emma               & 29           & F             & low vision      & since birth         & R: 20/2600, L: 20/2400 & R: none, L: limited              & guide dog              \\
Lily               & 42           & F             & blind           & since birth         & R: none, L: none       & R: very limited, L: none         & guide dog              \\
Zoey               & 53           & F             & low vision      & since birth         & R: none, L: 20/2300    & R: none, L: very limited         & guide dog              \\
Jing               & 34           & F             & low vision      & since 15 years old  & R: 20/200, L: none     & R: very limited, L: none         & white cane             \\
Jack               & 50           & M             & blind           & since birth         & R: none, L: none       & R: none, L: none                 & white cane             \\
Hanu               & 54           & F             & blind           & since child years   & R: none, L: none       & R: none, L: none                 & guide dog              \\
Amar               & 32           & M             & blind~          & since 21 years old  & R: none, L: none       & R: none, L: none                 & white cane             \\
Nora               & 48           & F             & low vision      & since birth         & R: unknown, L: unknown & R: limited, L: very limited      & white cane             \\
John               & 32           & M             & blind           & since birth         & R: none, L: none       & R: none, L: none                 & white cane             \\
Luca               & 35           & M             & low vision      & since birth         & R: 20/120, L: 20/120   & R: full, L: full                 & white cane             \\
Ryan               & 61           & M             & low vision      & since 12 years old  & R: 27/800, L: 27/800   & R: full, L: full                 & white cane             \\
Owen               & 44           & M             & low vision      & since teenage years & R: 20/2400, L: 20/2400 & R: very limited, L: very limited & white cane             \\
Yuan               & 33           & M             & blind           & since birth         & R: none, L: none       & R: none, L: none                 & white cane             \\
Noah               & 48           & M             & blind           & since 4 years old   & R: none, L: none       & R: none, L: none                 & guide dog              \\
Iven               & 40           & M             & low vision      & since 20 years old  & R: none, L: 2500       & R: none, L: very limited         & white cane             \\
Adam               & 59           & M             & blind           & since 40 years old  & R: none, L: none       & R: none, L: none                 & white cane             
\end{tblr}
\caption{Participant demographics. F=Female; M=Male; Visual acuity and visual field measures were self-reported by participants.}
\label{tab: Participants’ Demographics Information.}
\end{table*}

\subsection{Virtual Environments} \label{Virtual Environments}

We created four different virtual environments including the functions described above: a tutorial scene where participants could familiarize themselves with the controls and three task environments modeled after parks (Figure \ref{fig: Screenshots of the Tutorial and Three Park Environments from Our Prototype.}). We chose different features for each park to encourage exploration and observe participants’ ability to understand content using the guide. We also added environmental audio, such as recorded conversations for social avatars and general forest ambiance. 

The tutorial scene was a simple white room containing a purple table with two balls on top. Participants could move around and grab the balls to experiment with the study controls. This was designed to reflect a solitary onboarding space in most social VR applications.

The second scene was a brightly-lit medium-sized park (approximately 25 square meters) with two fountains, a river, and stone and wooden walkways crossing the river and around the park. Three groups of agent-avatars with pre-recorded conversations were distributed around the scene. 

The third scene was a dimly-lit large park (approximately 36 square meters) with three fountains, picnic tables and benches, and stone and grass walkways. Three groups of agent-avatars with pre-recorded conversations were scattered around the scene, along with five pieces of trash.

The fourth scene was a brightly-lit smaller park (approximately 16 square meters) with only a few plots of grass, stone walkways, and four groups of social avatars with pre-recorded conversations spread around the scene. One agent-avatar group was having a “dance party,” with music playing. 

\subsection{Participants and Recruitment} \label{Participants and Recruitment}

We recruited 16 participants with visual disabilities (10 male, 6 female) whose ages ranged from 29 to 61 (mean = 43.38, standard deviation (SD) = 10.30, Table 1). Participants were recruited through mailing lists. Participants were eligible for our study if they identified as blind or low vision, were at least 18 years old, could travel for an in-person study, and met the Meta Quest Health and Safety Guidelines \cite{meta-health-and-safety}. All procedures were approved by the Institutional Review Board at our university and all participants gave oral informed consent. We made a screening call to each participant to check their eligibility. During the screening call, we also asked about participants’ demographic information, VR experience, and use of mobility aids and visual interpretation services. Among the 16 participants, nine had VR experiences and 13 used visual interpretation services. Participants were compensated \$50 for participating in the study session and up to \$30 of compensation for travel expenditures upon presentation of a receipt. 

\begin{figure*}
\includegraphics[width=325pt]{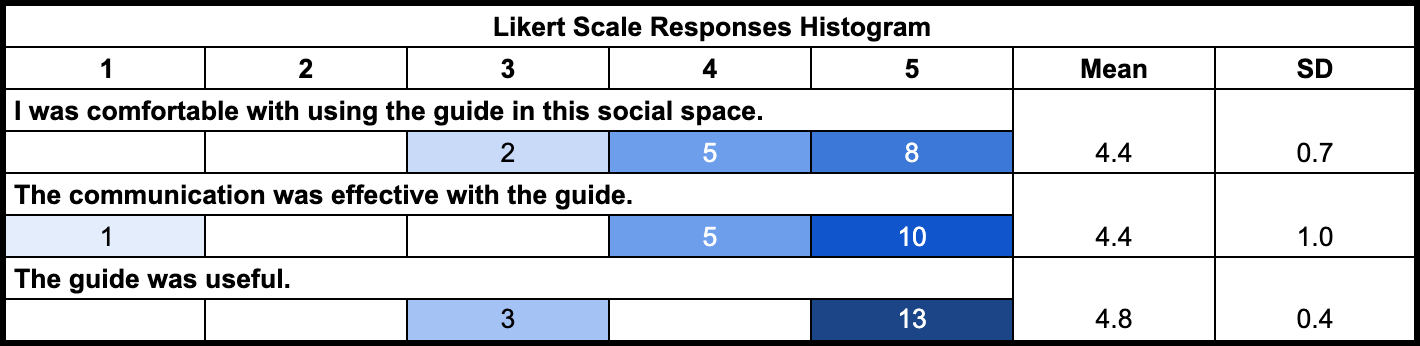}
\caption{\label{fig: Likert Scale Responses Histogram.} Likert Scale Responses Histogram. 1 = Strongly Disagree, 2 = Disagree, 3 = Neither Agree nor Disagree, 4 = Agree, 5 = Strongly Agree}
\end{figure*}

\subsection{Procedure}

The study was conducted in two locations, at a hotel during a conference and a university campus, and took about 1.5 hours per participant. Two researchers participated in the study: one acting as a guide in a remote setting and the other facilitating the study in-person with the participant. Both the participant and the guide used a Meta Quest 2 headset and controllers. After the tutorial, participants completed the three VR tasks with the guide, taking a break between each task. After each task, we asked participants to describe the park’s layout and contents. We determined that participants had an accurate understanding of the park if they were able to correctly describe key objects and the layout of the park. We ended the study with a 30-minute interview.

\textbf{Tutorial: VR System Introduction.} During the ten-minute tutorial, we explained (1) how to wear a VR headset and hold controllers, (2) the buttons on the controllers relevant to the study, and (3) the sound effects attached to the actions they would perform. Participants familiarized themselves with these features by moving around the virtual environment and interacting with objects within it. A guide was not present in this scene, as it was meant for the participants to get a feel for the controls on their own. The tutorial ended when participants had tried each control and were comfortable with them.
Next, we introduced the idea of a guide in VR that could aid the participants. We told participants they could communicate with their guide and grab onto their guide’s avatar to move with her. We explained she would not act on her own and would only act according to the participant’s instructions.

\textbf{Task 1: Free Exploration.} Participants had ten minutes to freely explore park one with the guide by moving around and interacting with objects. The idea for this first task was for the participant to gain a basic understanding of what the park was like and what objects were inside it. We observed how the participants became familiar with the guide and worked with them to navigate and explore the virtual environment. We did not add any other objectives to this task. 

\textbf{Task 2: Scavenger Hunt.} Next, participants had fifteen minutes to complete a scavenger hunt to find five pieces of trash in park two. If the participant “collected” (picked up) a piece of trash, the trash would disappear. We chose this task to encourage participants to be more involved and interact with the virtual environment. We again observed participant-guide interactions, and also how they used the guide for more challenging and specific tasks. 

\textbf{Task 3: Free Exploration with a Flying Guide.} In this task, participants were asked to imagine their guide as a bird or some other flying creature that could view the entire park from above. Participants were told that they could move around freely or with assistance, just as they did in previous parks. However, they could also fly with their guide, or ask the guide to fly around and return to them. Participants had ten minutes to explore park three and their guide’s flight ability. We again observed participant-guide interactions, expecting participants would now be familiar with the guide and observing whether their use of the guide changed as a result. In addition, the “flying guide” was meant to demonstrate the absence of the usual physical constraints of physical guides. We hoped this would encourage them to think more imaginatively about how a guide in VR could support their needs.

\textbf{Post-Task Interview.} We ended the study with a 30-minute interview, in which we asked participants to reflect on their experience and discuss possible improvements, including their thoughts about the guide’s appearance, functions, and methods of communicating with the guide. Participants also gave responses to three Likert scale statements regarding the comfort, effectiveness, and usefulness of the guide interactions.

\subsection{Data and Analysis} \label{Data and Analysis}

All participants completed all three tasks, except for one participant who did not attempt the third task due to physical discomfort. All participants completed the post-task interview. Audio and video recordings from the study session were collected from the guide’s headset. We transcribed the sessions using Otter.ai, an automatic transcription service. Two researchers coded the transcripts using open descriptive codes. We began by coding the same three transcripts, then came together and discussed discrepancies. Through the discussion, we generated a codebook and split the rest of the data. Afterward, we conducted a thematic analysis \cite{braun_using_2008} using affinity diagrams to categorize the codes into themes.

\section{Findings} \label{Findings}
\subsection{Overview} \label{Overview}

\begin{figure*}
    \centering
    \begin{subfigure}[b]{0.49\textwidth}
      \centering
      \includegraphics[width=\textwidth] {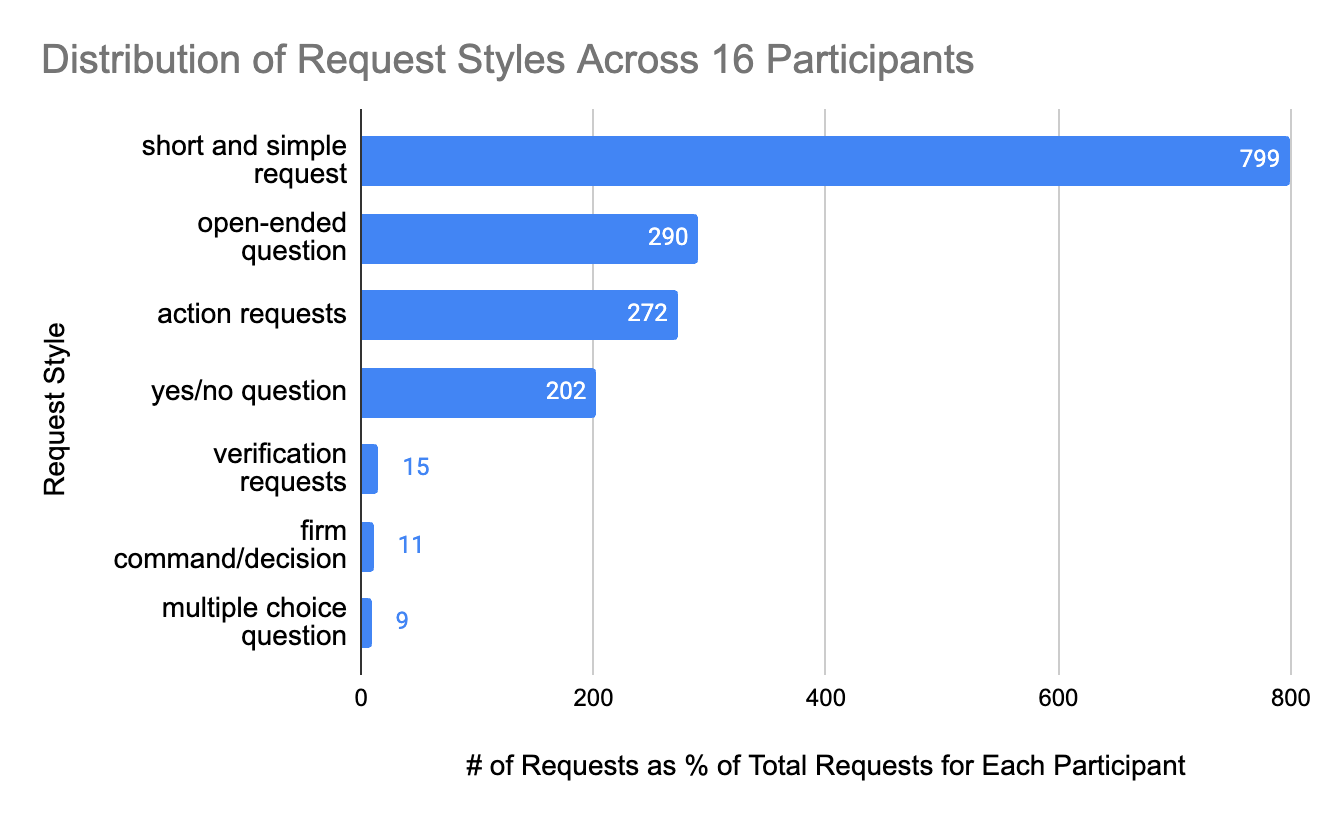}
      \caption{Request Styles Across 16 Participants.}
      \label{fig: Request Styles}
    \end{subfigure}
    \hfill
    \begin{subfigure}[b]{0.49\textwidth}
      \centering
      \includegraphics[width=\textwidth] {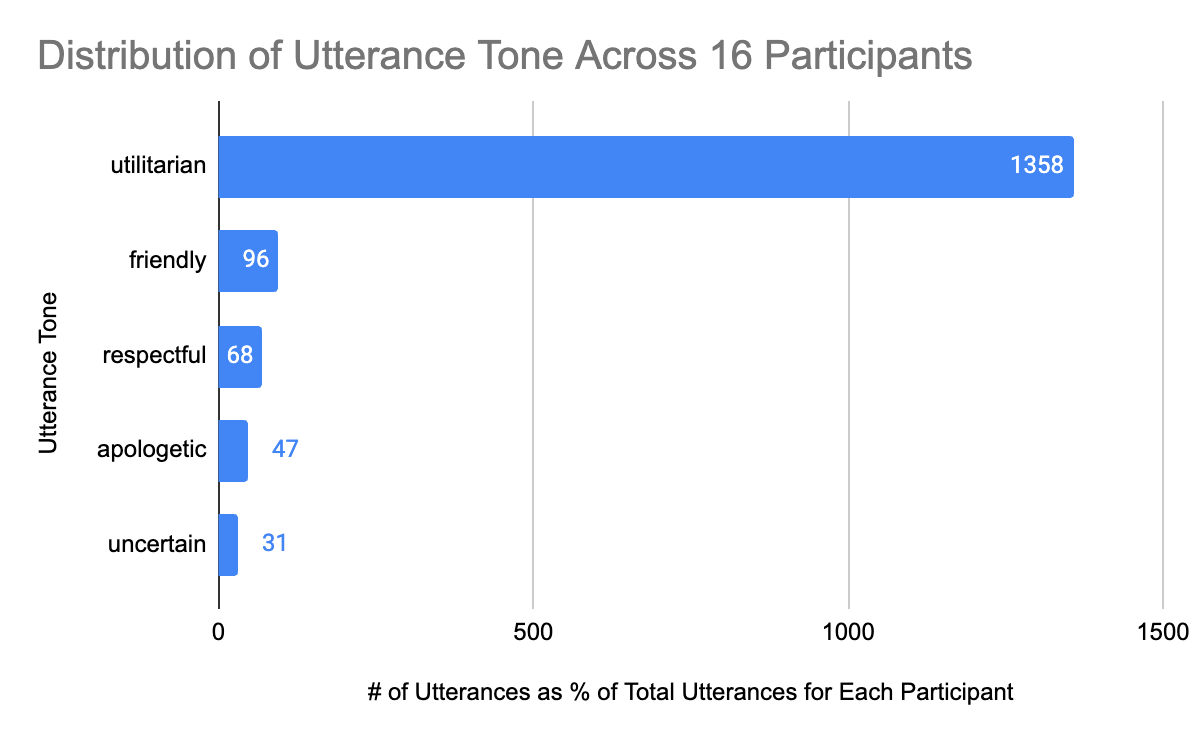}
      \caption{Utterance Tone Across 16 Participants.}
      \label{fig: Utterance Tone}
    \end{subfigure}
\caption{Characteristics of participant requests and utterances to the guide. They are normalized as the percent of the participant’s total number of requests or utterances to avoid biasing the data toward talkative participants. Left: requests fell into seven style categories seen above. Right: utterance tones included (1) utilitarian, the request purely fulfilled a need; (2) friendly, a pleasant or humorous statement; (3) respectful, polite expressions; (4) apologetic, participants apologized for actions or outcomes; (5) uncertain, they deferred to the guide or gave them authority.}
\label{fig: Request Charts Combined}
\end{figure*}

All participants were able to explore the virtual environments and a third of the participants experimented with teleportation for movement with varying levels of success. During the scavenger hunt (Task 2), nine participants found all five pieces of trash, one found four pieces, five found three pieces, and one found one piece. During Task 3, 12 participants chose to use the guide’s flying feature. All participants accurately described the parks with varying levels of length and descriptiveness.

Figure \ref{fig: Likert Scale Responses Histogram.} shows participant responses to Likert scale statements. One participant did not specify a number for the first statement, explaining that using a guide was “natural” and was not applicable to “comfort” (Hanu).

Overall, participants’ scores demonstrated that they felt comfortable using the guide in a social space (mean = 4.4, SD = 0.7), found communication effective (mean = 4.4, SD = 1.0), and found the guide to be useful (mean = 4.8, SD = 0.4). Most participants found the guide helpful without it being intrusive. Noah explained that the guide “provided information when requested” and “waited for me to suggest what it is that I wanted to do.” Participants also felt communication was effective, sharing that “communication was great. I was comfortable asking [the guide] questions” (Jing). Interestingly, one participant (Hanu) rated communication a “1, strongly disagree.” Hanu explained that the guide should “offer more information, options, and choices.” Lastly, most participants strongly agreed that the guide was useful, stating the guide “was very useful because I needed more help than I thought I needed” (Jing) and that “you could get a lot of information” (Lily). 

\subsection{Participant Interactions with the Guide} \label{Participant Interactions with the Guide}

Participants requested different kinds of information from the guide to fulfill accessibility needs. We categorized these requests into asking about objects, the virtual environment, and technical controls.

Participants’ requests about objects served several purposes: they asked about an object’s appearance, ways in which they could interact with it, or object locations. Examples of these requests, from most to least common, included:

\begin{itemize}
\item Asking where an object is located in relation to one’s body (Adam: “From where I am now, where would you say that is?”)
\item Asking if the guide sees a particular kind of object (Hanu: “Are there any picnic tables or anything around us?”)
\item Asking about people in the virtual environment (Owen: “Is there a person over there?”)
\item Asking a specific question about the virtual environment (Amar: “Is there anything else around this fountain?”)
\item Asking what an object looks like (Amar: “What type of fountain is this?”)
\end{itemize}

Participants’ questions about the virtual environment included requests for contextual information, new or unexplored areas, and the time of day in the environment. Examples of these requests, from most to least common, included:

\begin{itemize}
\item Asking what the participant is looking at (Jing: “What am I looking at now?”)
\item Asking what obstacle they ran into (Noah: “Am I running into the gazebo here?”)
\item Asking for contextual information about the virtual environment (Jack: “What’s going on [in this virtual environment]?”)
\end{itemize}

Interestingly, even though participants knew there were no other “real” people in the environment besides the guide, they asked a range of questions about the agent-avatars present, from their actions to their relationships. For instance, Jack asked, "What are people generally doing? Are they walking around, sitting around, or having picnics?" Jack was particularly interested in the relationship between individuals in the environment, asking, "[are they] a couple?" Similarly, Amar probed for the identities of people around him: "Who are the people on the bridge?” 
Participants also asked which controls they should use to perform a specific action. Examples of these requests, from most to least common, included:

\begin{itemize}
\item Asking about controls in VR (Ryan: “How do I do the snap turns?”)
\item Asking how to perform a specific action (Jing: “Am I only supposed to hit the grab trigger?”)
\item Asking where to aim to teleport (John: “Okay, so where do I aim now? Straight?”)
\end{itemize}

Participants also frequently asked the guide to verify their understanding of the environment, including their location, needed movements (i.e., were they doing or about to do the correct action), and control use (i.e., were they using or about to use the right control). For example, Ryan asked, “So we’re in the gazebo, right?”

In addition to requests for information, participants asked the guide to perform actions. They often asked the guide to interact with objects, such as, “Can you pick it up and hand it to me?” (Jack). During Shared Movement, participants asked the guide to move them in specific directions, such as “Can we continue on a path? Like is there a path that goes all the way around it?” (Lily) This type of movement request was one of the most common. These differing strategies demonstrated how the guide supported varied needs in the virtual environment. 

Participants posed different question types or requests to the guide. We present this distribution of “request styles” in Figure \ref{fig: Request Styles}. For example, most participants’ requests were short and simple (Luca: “Yeah, how do I walk forward?“) and were posed as open-ended (Ryan: “What is this green thing?”) or yes or no questions (Owen: “Is the cup on the other side of the fountain?”). 

Beyond different styles of requests, participants also had different utterance tones with differing underlying manners or emotions. Utterances included all statements made by the participants including requests. We present the spread of “tones” in Figure \ref{fig: Utterance Tone}. Most utterances were utilitarian when communicating with the guide (Owen: “Are we getting close?”), whereas others were friendly (John: “How are you? Pleasure to meet you.”), respectful (Yuan: “Okay, there we are, thank you.”), or uncertain (Nora: “So what are we going to do now?”).

\subsection{Differences Between Blind and Low Vision Participants}

We noted distinct differences between how blind and low vision participants utilized the guide. As one might expect, blind participants asked for more visual interpretation from the guide than our low vision participants, asking the guide for descriptions of objects, people, and the virtual environment. Meanwhile, low vision participants often described what they saw and sought the guide’s verification, rather than asking the guide to describe visual information directly.

Blind and low vision participants demonstrated distinct patterns in movement as well. In general, low vision participants used more movement controls (walking, teleportation, and flying) throughout their experience. All eight low vision participants experimented with walking, Shared Movement, teleportation, and flying at some point during their experience. Low vision participants also often explored independently of the guide. Six low vision participants moved primarily by walking independently, as opposed to using Shared Movement with the guide (the other two participants used Shared Movement). These participants used Shared Movement when getting familiar with the VR experience at first, but used it thereafter when encountering difficult visual tasks. For example, Jing, who had low vision, engaged in Shared Movement when trying to pick up a brick from the top of a trash can. Since the brick’s color did not contrast with the trash can, she could not position herself properly to grab it.

In contrast, blind participants tended to use Shared Movement throughout their entire VR experience, often asking for guidance as soon as they entered the parks. Out of eight blind participants, five used Shared Movement with the guide as their primary movement method (the other three participants used the walking control). Surprisingly, two of the blind participants did not use Shared Movement at all. This was part of another trend we observed: blind participants did not experiment as much as low vision participants with the various controls, typically picking one control they preferred and sticking to it. One of our blind participants, Lily, explained this inclination, stating, “I probably should have used the joystick more to like, kind of explore it more independently. But it was nice to have the guide to ask [for help].” Only three blind participants tried all four movement controls (walking, Shared Movement, teleportation, and flying) at some point during their experience. In addition, four blind participants did not try flying, even after it was introduced in the third park. 

One interesting distinction between blind and low vision participants was how they used the teleportation control. Only two low vision participants out of eight tried this control, whereas five blind participants out of eight tried it. The blind participants used the teleportation control in unexpected ways, like Yuan, who used it to try and climb a tree. 

\subsection{Reflections on Using a Guide} \label{Reflections on Using a Guide}

Participants perceived the guide in three primary ways: as a tool, a companion, or a superfluous presence that hindered their independence.

For nine participants that viewed the guide as a tool, we observed the guide was mainly used for utilitarian queries for information or requests for help with some action. Over 90\% of their requests were identified as utilitarian. These participants did not ask personal questions, interact with the guide aside from their accessibility needs, or use humorous or casual language. Importantly, though, this did not mean the participant did not enjoy using the guide.

In contrast, five participants viewed the guide as a companion and tried to include her in their experience. Unlike participants who viewed the guide as a tool, this group of participants made a significant number of non-utilitarian requests. Twenty to 55\% of their requests fell into categories such as friendly, respectful, apologetic, or uncertain (in order of frequency). They called the guide by her name instead of just “guide,” and treated the guide like a friend. For example, John asked the guide personal questions, such as, “By the way, Easter's coming up. Did you ever do scavenger hunts?” He then engaged the guide in a conversation about TV shows from his childhood. When John and the guide approached a group of dancing agent-avatars, John encouraged the guide to dance with them: “Oh hey, hey, hey, hey! That's something we're doing. You should try it out!” 

Other participants felt the guide supported and empowered them. Nora referred to the guide as a “friend,” saying the guide “[made] me stronger.” Luca shared a similar sentiment on the guide’s companionship, stating that “the guide has your back.” 

While most participants felt supported by the guide, four felt the guide hindered their independence (two of these four were among those who viewed the guide as a tool). For example, Amar stated, “The [guidance] experience made me feel more dependent than I needed.” Ryan echoed this sentiment, saying that “guiding implies dependence.” It seemed participants who felt dependent on the guide considered this a negative aspect of their experience. Noah was another participant who felt this way, and recommended replacing the human guide with an AI-powered one:

\begin{quote}
    I don't think there's anything that the guide could do to make [the experience] feel more independent, other than if we were able to replace the human guide with an AI guide. I am not relying on a person, I am relying on software that I am telling what to do. That's the difference between asking you to read the computer screen and me using the screen reader to read the computer screen.
\end{quote}

\subsection{Reflections on the Guide’s Initiative} \label{Reflections on the Guide’s Initiative}

We now discuss possible approaches to designing VR guides based on participants’ reflections.

Participants wanted varying levels of initiative from the guide to have a supportive companion that fulfilled their needs. Hanu felt the guide’s lack of initiative lowered her trust in her competence; she described it as:

\begin{quote}
    The guide was not making me trust them. They seemed to be waiting too much for instruction for me, and they didn't seem to have the confidence that I would trust that they would be doing the right thing… I think, especially when they're in the guide mode, they should be in control.
\end{quote}

On the other hand, several participants appreciated the guide’s somewhat passive approach to guiding. Yuan mentioned it was nice that the guide was not “overly helpful” and tried to “guide me under my guidance,” rather than assuming she knew how to help. Lily agreed, saying the guide did well by asking “‘How can I help you in this situation?’ instead of just doing something for me without asking.”

Most participants wanted the guide to provide more preemptive information in certain circumstances. One example was giving information if it was urgent or necessary. For example, Noah said he wanted a notice: “When I’m about to walk into the bushes.” Other participants agreed the guide should alert them of obstacles in advance, whether they asked for it or not. Noah also suggested that “inter-human information” might be important, such as when a group of people suddenly begins staring at or coming towards them. 

\subsection{Reflections on the Guide’s Functions} \label{Reflections on the Guide’s Functions}

While some participants focused on the interpersonal interactions they had with the guide, others brought up the guide’s supportive functions.

Many participants highlighted issues relating to orientation and navigation. While they found the guide helpful for these tasks, they offered various potential improvements. Jack reflected on his difficulty using the VR controller to pick up trash during Task 2, and said it would be great if the guide could “help you get the joystick pointed at the object.” This might be a function used only when participants were getting used to VR, though, as Jack added “You might want to learn how to do it yourself too.” Supporting this, Ryan described his preferences for the guide’s navigation assistance as:

\begin{quote}
    I preferred actually just having the guide as a standby and doing the mobility stuff myself rather than being led, but the guide was useful for, if I needed, clarification, validation, or whatever. So it was kind of like, just to have the guide in my back pocket, rather than leading me.
\end{quote}

As participants grew more comfortable with VR controls, they also seemed more comfortable minimizing the guide’s role in navigation. Conversely, while they were in an unfamiliar environment or still learning VR controls, they thought giving the guide more control over movement would be helpful.

During the scavenger hunt, participants were satisfied with the minimal functions the guide had (Adam). Namely, they liked the guide’s ability to fetch objects and said that “fetching objects is a good [function]” (Emma) to keep for the future. 

In addition to providing information and performing actions, participants wanted the guide to modify environments directly to make them more accessible. Participants gave examples of inaccessible environments like “chaotic” (Iven) spaces that are missing accessible audio feedback or are cluttered with obstacles. Yuan suggested the guide could “zoom in” on a particular audio source if there were excessive overlapping sounds. Jing suggested the guide could “change the color” of objects to support her low vision. Several participants also suggested adding sound beacons or haptic “buzz areas” on objects to notify them that they were getting close (John, Ryan, Noah, Iven). In general, the idea of altering a confusing environment to suit the participant’s preferences seemed like a desirable function.

Participants also wanted support in social situations. For example, they wanted the guide to describe people’s appearances or determine people’s approachability. Amar was interested in identifying coworkers at virtual work functions, and imagined the guide helping him find a particular person: “I would say, that gentleman there? Does he have long brown wavy hair? Is he wearing glasses?” He later added to his scenario, saying it would be nice if the guide could tell him “if someone will talk to you.” Noah imagined a similar use for the guide, where if he overheard an interesting conversation, the guide could lead him “right next to [the conversation],” so he could join easily. He described an ideal interaction as:

\begin{quote}
    I could walk in there and have [the guide say], ‘Okay, here's the layout of the room. I don't see Joe that you wanted to talk to, but I do see these other people who, they're working on this project, maybe you want to go talk to them.’ When I go to a party, like a work party to hang out…there are people that I know who I want to talk to, and maybe there are some people who I don't know that I want to talk to because they seem interesting.
\end{quote}

Participants had varying opinions on flight and other “fantasy” abilities enabled by VR. Amar mentioned that non-natural abilities like flight wouldn’t be useful unless he was in some kind of fantasy experience. In contrast, other participants were more inclined to accept non-natural abilities due to the nature of VR. As Noah put it when describing teleportation, “it's VR, you enter a bit of randomness” and accept your strange abilities. Jack added that abilities like flight are “quicker [than traditional methods] to get everything from up above” and learn what a scene is like.

\subsection{Reflections on the Guide’s Form and Visibility} \label{Reflections on the Guide’s Form and Visibility}

Participants were divided in how much they cared about the guide’s appearance. Some thought it was very important, as it might leave a positive or negative impression on people around them, but others said it would not matter at all. These participants generally agreed that since the point of the guide was “the information they’re giving” (Zoey), it should not really matter what they looked like (Zoey, Jing).

Participants who cared about the guide’s appearance had a range of preferences. Most wanted the appearance to be “situational” (Owen), decided on a case-by-case basis. These situations were generally dependent on formal or informal settings, such as whether they were playing a game or attending a “virtual conference” (Ryan). Luca compared it to a video game where you could “choose which character you want to play with” in case there are “aesthetics” you want the guide to embody. However, these “aesthetics” varied greatly.

Some thought the guide should look like part of the participant’s own avatar. For example, the guide could appear as a machine, an assistive technology device, or even a small animal, like an accessory of their avatar. Lily explained this design choice by describing how she “didn't really envision [the guide] as an individual” and thought of them as an assistive tool rather than a companion. Yuan added it would be best if the guide could “melt into the background or maybe be thought of as part of the person's image…rather than recognize the fact that it's an individual in and of itself” to bolster his independence.

Yuan was one of several participants interested in the guide being a discreet bird sitting on their shoulders. In fact, Ryan, Owen, Yuan, and Noah all mentioned one ideal appearance for the guide would be a “parrot on my shoulder,” offering guidance unobtrusively so the act of being guided was not obvious. Yuan further elaborated on his ideal “parrot,” saying he would want to customize the bird based on his personal interests:

\begin{quote}
    I have a great interest in fantasy and mythology. So if I were to make some sort of avatar, it would probably have some sort of fantasy basis…most likely that avatar would have a hawk perching on its shoulder or maybe as I said, two ravens perching on my shoulder because I just think it would look like nice, so it would probably be part of my image anyway. Anyone who knows me–assuming that some of the people there know me–would probably just say, ‘Oh, okay, he's with two ravens perched on his shoulder. I can see [Yuan] doing that.’
\end{quote}

Other participants wanted the guide to appear as another individual. These participants preferred human guides to other forms and wanted to build a human connection with the guide. John said he liked the human guide because it made the experience “more like home,” as though he could “imagine walking with somebody instead of a creature.” The idea of being comforted by another person’s presence was echoed by multiple participants, including Nora who liked that “I have somebody there like myself.” Both Amar and Nora also preferred a guide whose appearance reflected the guidance experiences they had in real-life. Nora and Amar had never used guide dogs, so they felt “more natural interacting with a human” (Amar). One of the most interesting perspectives came from Hanu, who tied her preference for a human guide to cultural perceptions in the disability community:

\begin{quote}
    If the person you're working with is Deaf, there are boundaries such as that you don't speak to the interpreter, you speak to the [Deaf] person. But when I'm working with a guide, you don't ignore the guide. To the other [blind] people in the group, the guide becomes part of the group. I think that's just cultural differences.
\end{quote}

While the guide’s appearance was important to participants, we asked whether it needed an appearance to begin with. Participants discussed the guide’s visibility: whether or not it could be seen by anyone else. Some had strong opinions that the guide should always be invisible or visible while others did not care. For instance, visibility did not matter at all to Lily and Jing. Lily only saw possible “fun” in using an invisible guide to impress other people, “because people would be like, ‘It’s so amazing. You’re blind, but you’re still doing all this [without a guide].’” In contrast, Owen believed that having an invisible guide would be “the ultimate experience” because it would mean “nobody would know I was using the guide.” Yuan agreed that the invisible guide would create a better experience, saying he would be “a lot more comfortable with the idea” of guiding if no one else knew about the guide.

We found participants who preferred a visible guide believed the guide’s presence would give a more accurate impression of their needs. Nora compared it to using a white cane to signal her disability: “Once I have the cane out, they know [and] they’re gonna help.” Emma and John expanded on this idea of “showing” people the guide, saying that a visible guide makes their presence known. Additionally, John believed a visible guide would be especially good for those who are “newly blind,” as it would show them the BLV community exists in VR and “make them feel comfortable that yes, there is help out there.” For these participants, the guide’s visibility was a source of comfort or empowerment, a way of helping the BLV community demonstrate their presence in VR.

As with the guide’s appearance, participants said visibility should also change depending on the situation. Once again, the best practice seems to be allowing for customization across formal or informal scenarios in order to create the best guide possible.

\subsection{Future Possibilities: An AI Guide} \label{Future Possibilities: An AI Guide}

As seen above, participants made various suggestions for guide improvements based on what they experienced. We also asked participants to think beyond the human guide format they experienced and consider how they would feel about an AI guide instead.

Participants reacted positively to a potential AI guide, saying it afforded more independence than a human. Noah and Ryan gave similar sentiments that the AI would make them feel like less of a “burden” (Ryan) since it wouldn’t be a sighted person who had to “drag [them] around” (Noah) but a tool designed for the purpose of guiding. Additionally, Amar pointed out that the AI guide is “instant” and always available, so he wouldn’t have to worry about whether a sighted person had time to help. AI’s convenience and immediacy made it an attractive guidance option.

Even still, many participants–including those who would consider AI forms–had reservations about employing AI for visual guidance. These reservations boiled down to issues of trust and reliability, or preferences for human interaction. Though Amar liked the AI’s convenience, he worried about its performance: “I don’t want to join a social setting and we’re walking and the AI has an issue and I end up in the fountain.” Owen simply said he finds AI to be “way less effective than a human, particularly when it comes to describing just visual things.” Iven suggested that even if you had an AI guide, you should still “have a human as a backup,” in case something goes wrong. Finally, John pointed out that whether the AI functioned well or not, “I personally prefer the human guide because of the social interaction.”

\section{Discussion} \label{Discussion}

To our knowledge, this paper presents the first exploration of using sighted guides to enhance the accessibility of VR for the BLV community. Our basic guide framework enabled us to observe when and how participants interacted with the guide. Moreover, it provided participants a concrete experience, upon which they could reflect and imagine future possibilities. As such, the study allowed us to identify novel participant needs in VR and opportunities for the design of virtual guidance systems. These needs and opportunities, along with our observations of granular participant-guide interactions (section 4.2), represent the main contributions of our work. Our basic guide framework (including guidance scaffolds and behavior guidelines) represents a secondary contribution, as it can serve as a baseline for future guidance systems. 

\subsection{Identifying Novel Needs in VR for BLV Users} \label{Identifying Novel Needs in VR for BLV Users}

	Beyond navigation and environment and object perception, prior work in social VR and virtual worlds has uncovered certain accessibility needs for BLV people. These include the need for avatar representation for disabilities \cite{zhang-part-of-me-avatar-diversity}, detecting avatar proximity \cite{ji-vrbubble}, and creating accessible communication and technical controls \cite{khalifa-radegast, oktay-textsl-secondlife, trewin-powerup}.

Adding to these needs, our study uncovered novel accessibility concerns in VR. First, we discovered the need to support embodied social interaction, such as proximity and “natural” movement. For example, participants wanted the guide’s help in moving towards social avatars and joining conversations in “natural” ways, rather than moving haphazardly or bumping into objects along the way, which could highlight their disability or make them seem incompetent.

Next, we noted a subgroup of participants benefited from social support in the VR experience. The guide’s role as a companion was key to bolstering their confidence in an unfamiliar experience or feeling comfortable interacting with the virtual environment (i.e., assuring them that nothing was going wrong). This indicates a need for a designated companion for the user in the social space. Although this may not be a unique need for BLV users, we must think of them as people first, rather than defined by their disability-specific needs. 

We also found a need for user control over inaccessible environments. Many of the participants are used to addressing daily accessibility problems by directly modifying the world around them. For example, adding Braille labels to kitchen appliances \cite{wisconsin-council-household-products-labeling} or increasing the lighting in their houses \cite{boston-sight-safe-home}. In our study, this manifested through participants’ interest in the guide altering aspects of the environment to make it more accessible (section 4.5). 

Participants also wanted contextual information about the people, objects, and virtual environment. Participants wanted to learn the limitations of possible actions in the virtual space and the “history” behind certain features. For example, understanding why certain objects were included (e.g., why was there a fountain in the path?) or the guide’s opinion on why social actions were occurring (e.g., why were people dancing?). This information can be inferred by contextual cues, but to engage with a space, participants wanted more than typical superficial descriptions (e.g., alt text). 

Lastly, participants needed support becoming familiar with novel VR controls. VR controls have not worked their way into mainstream devices, and remain unfamiliar to most users. While prior work \cite{khalifa-radegast, oktay-textsl-secondlife} has focused on using daily technology such as keyboards to create access to non-immersive virtual worlds (e.g., SecondLife), immersive VR requires users to adapt to novel technologies like VR controllers in order to achieve that access. As more advanced VR controls are created, we identify a need for continuous support with such controls over time, similar to a helpline that can be used as a frequent reference or troubleshooting service. Importantly, BLV people do not benefit from incidental learning that sighted people experience by seeing the use of devices, so they often need explicit support. 

Building on novel and known needs, there are many possible roles a guide can fulfill beyond basic navigation and scene understanding support. These needs should be considered going forward by the VR accessibility community through other approaches as well.

\subsection{Opportunities for the Design of Virtual Guidance Systems} \label{Opportunities for the Design of Virtual Guidance Systems}

Our exploratory study unveiled a massive design space for future guidance systems. We outline key opportunities below.

Virtual guides could be powered by AI to build user independence, as participants suggested. These AI guides could use computer vision models or built-in descriptions attached to objects or environments to relay information \cite{zhao2019seeingvr, zhao-cue-see, winlock-realtime-grocery-detection}. They could work similarly to voice assistants with a voice-based question and answer system \cite{fingas-how-does-alexa-work}. AI guides might also be designed to act proactively, offering immediate information based on the user’s preferences. For example, a user could have an AI guide describe the environment’s layout, then offer further detail. AI guides could even combine these styles of offering proactive information and waiting for user input. There is room for a middle ground prior to AI guides by developing AI-powered screen readers instead. In fact, researchers have already explored creating VR screen readers \cite{sun2023alt}, through applying alt-text and other sensory modalities such as haptics to digital objects. This approach seeks to enforce existing standards for web accessibility in VR and allow BLV users to make use of technology they are already familiar with, such as screen readers. However, even on the web, accessibility standards are poorly followed and do not always address users’ access needs \cite{power2012guidelines}. An AI guide could use whatever alt-text or sensory information exists in the VR scene, and further add to or clarify that information for the BLV user to make sure access needs are met. More importantly, an AI guide would be a unique accessibility tool that is embodied, conversational, and present with the user in VR.

Different human guide frameworks warrant exploration. Since the virtual guide is embodied in the inaccessible scene alongside the BLV user, it provides social as well as descriptive support, creating a new field of embodied remote assistance. However, we can still find inspiration for guide frameworks from crowdsourcing and “friendsourcing” for alt-text image descriptions on social media, connecting BLV users to remote sighted help \cite{brady-visual-challenges-everyday-lives, salisbury-conversational-crowdsourcing}. One example framework could be a volunteer system that matches sighted individuals on social VR platforms with BLV users who request guides. The system could create matches by using data on the VR experiences both users enjoy to pair users with their ideal guide. A “friends-based” system could also connect BLV users to people on their friends lists if they are online. Such a guide system could be commercial, like the visual interpreting service, Aira, with a company hiring and training guides that offer professional assistance to BLV customers. It should be noted that while both friend-sourcing and crowd-sourcing can be considered, participants were largely reluctant to rely on figures they considered friends or family members to meet their access needs. They generally preferred to conceive the guide as a tool, to remove the notion of “burdening” their social network with said needs. Thus, a crowd-sourcing framework might be more practical to pursue initially given participant views.
  
Guide training is another area for development. Notably, training might go beyond current standards for remote-sighted help, since VR is an embodied medium that warrants consideration of the guide’s tone of voice and body language. Even so, visual interpretation services (e.g., Aira, BeMyEyes) offer training materials for sighted workers that could inform this topic \cite{be-my-eyes-sighted-volunteer, aira-agent}. We introduced a sighted guide with “naive” training (section 3.2) which worked well for most participants, but not all. Future training, or AI programming, might include tips on what information guides should offer, how guides should phrase information, or how guides should make suggestions without undermining the user’s independence. 

Without a doubt, training in body language would be most applicable if the guide’s avatar was human, thereby opening up for continued exploration. Virtual guides, whether they are human or AI, can have a huge variety of appearances in VR. Participants gave a range of preferences for these appearances, from birds perched on their shoulders to average humans walking beside them, and mentioned how both appearance and visibility affect the quality of their VR experiences. Which form is the most helpful for BLV users in VR? Are there situations that lend themselves to invisible or visible guides? Guide appearance is a rich area of inquiry for future work.

Our study included basic functions for the guide, but future guides could tap into many more VR-specific abilities; for example, addressing issues with teleportation (section 5.3) by temporarily manipulating the user’s controls. Participants proposed guide abilities to alter the environment by adding accessible audio or haptic feedback or altering visuals, supporting Balasubramanian et al.’s early look into dynamically altering inaccessible VR environments for BLV users \cite{balasubramanian-scene-weaver}. Guides could even collect a “sound library” (John) from a virtual environment so the user can familiarize themselves with sound effects before entering new virtual spaces. 

Our guide communicated through speech, but other feedback modalities warrant the exploration of supporting–or replacing–speech. VR presents the possibility of many novel communication methods. Guides could answer yes-or-no questions discreetly and quickly using haptic patterns. Frequent questions like basic environment descriptions could be translated into button shortcuts on a controller. Hand gestures could map to guide requests, or users could “pinch” and move the guide to virtual areas they are interested in.

We envision virtual guides to be a highly generalizable accessibility tool for VR, a framework that can be “dragged and dropped” into applications so guides can support users in any virtual environment. The features and forms of this guide may vary significantly between the platforms it joins, and there is no correct guide form we can point to now. However, our study has shown the potential for guides as a versatile assistive technology in VR that merit further exploration.

\subsection{Limitations} \label{Limitations}

One limitation of our study was that our virtual environment used computer-controlled agent-avatars who did not interact with participants, which might have changed the way participants used the guide for social interaction. The agent-avatars in our virtual environment were not acting in unpredictable ways, which may be the case in real social VR scenarios. For instance, participants may have asked more questions about real people in the environment or used the guide to interact with them. Future research in social spaces with real users can uncover different aspects of how guides can be used as well as possible social pressures around using them.

Secondly, although some participants had prior experiences with VR, most of these experiences were very short, such as in other research studies. More experience with VR would increase participant familiarity with controls and could also change the ways they used the guide. Some common controls in VR were difficult to use due to their visual nature. Teleportation was especially difficult because it relies on pointing to a visible teleportation target. Snap-turning was also challenging—participants would accidentally flick or hold the joystick and not realize their avatar had turned multiple times, even with sound effects attached to the movements. This exacerbated the role of technical support the guide had to play; the interaction patterns we observed may have been slightly different in an environment with a more accessible baseline.
\section{Conclusion} \label{Conclusion}

In this paper, we conducted a study exploring the use of a guide for BLV people in social VR with basic social VR functionality and additional accessibility scaffolds. Our observational study with 16 blind and low vision participants highlighted how participants used the guide and a range of participant preferences for guide initiative, functions, form, and visibility, pointing to a novel and exciting avenue of social VR accessibility research.
\section{Acknowledgements}
This paper is based upon work supported in part by the National Science Foundation under Grant No. 2212396, and a gift from Meta (Meta Platforms, Inc.). The contents of this paper do not necessarily represent the policy of the funders, and no endorsement should be assumed. We thank all the participants for their time. We also thank Mahika Phutane, Lucy Jiang, and Ricardo Gonzalez for their feedback on the study design and manuscript.

\bibliographystyle{plainnat}
\bibliography{refs}

\begin{thebibliography}{70}
\providecommand{\natexlab}[1]{#1}
\providecommand{\url}[1]{\texttt{#1}}
\expandafter\ifx\csname urlstyle\endcsname\relax
  \providecommand{\doi}[1]{doi: #1}\else
  \providecommand{\doi}{doi: \begingroup \urlstyle{rm}\Url}\fi

\bibitem[bli()]{blindswordsman}
Blind {Swordsman}.
\newblock URL \url{https://vrjam.devpost.com/submissions/36270}.

\bibitem[sig()]{sightedguide}
Sighted/{Human} {Guide}: {One} {Instructor}'s {Perspective}.
\newblock URL
  \url{https://nfb.org/sites/default/files/images/nfb/publications/fr/fr34/1/fr340110.htm}.

\bibitem[Aira({\natexlab{a}})]{aira-agent}
Aira.
\newblock Become an aira agent - aira, {\natexlab{a}}.
\newblock URL \url{https://aira.io/our-agents/}.

\bibitem[Aira({\natexlab{b}})]{aira-homepage}
Aira.
\newblock Aira: Visual interpreting – get live, on-demand access to visual
  information, {\natexlab{b}}.
\newblock URL \url{https://aira.io}.

\bibitem[Andrade et~al.(2018)Andrade, Baker, Waycott, and
  Vetere]{andrade-echo-house}
Ronny Andrade, Steven Baker, Jenny Waycott, and Frank Vetere.
\newblock Echo-house: Exploring a virtual environment by using echolocation.
\newblock OzCHI '18, page 278–289, New York, NY, USA, 2018. Association for
  Computing Machinery.
\newblock ISBN 9781450361880.
\newblock \doi{10.1145/3292147.3292163}.
\newblock URL \url{https://doi.org/10.1145/3292147.3292163}.

\bibitem[Avila et~al.(2016)Avila, Wolf, Brock, and
  Henze]{avila-survey-about-be-my-eyes}
Mauro Avila, Katrin Wolf, Anke Brock, and Niels Henze.
\newblock Remote assistance for blind users in daily life: A survey about be my
  eyes.
\newblock In \emph{Proceedings of the 9th ACM International Conference on
  PErvasive Technologies Related to Assistive Environments}, PETRA '16, New
  York, NY, USA, 2016. Association for Computing Machinery.
\newblock ISBN 9781450343374.
\newblock \doi{10.1145/2910674.2935839}.
\newblock URL \url{https://doi.org/10.1145/2910674.2935839}.

\bibitem[Bailenson et~al.(2004)Bailenson, Beall, Loomis, Blascovich, and
  Turk]{bailenson2004transformed}
Jeremy~N Bailenson, Andrew~C Beall, Jack Loomis, Jim Blascovich, and Matthew
  Turk.
\newblock Transformed social interaction: Decoupling representation from
  behavior and form in collaborative virtual environments.
\newblock \emph{Presence: Teleoperators \& Virtual Environments}, 13\penalty0
  (4):\penalty0 428--441, 2004.
\newblock URL \url{https://web.stanford.edu/~bailenso/papers/TSI.pdf}.

\bibitem[Balasubramanian et~al.(2023)Balasubramanian, Morrison, Grayson,
  Makhataeva, Marques, Gable, Perez, and Cutrell]{balasubramanian-scene-weaver}
Harshadha Balasubramanian, Cecily Morrison, Martin Grayson, Zhanat Makhataeva,
  Rita~Faia Marques, Thomas Gable, Dalya Perez, and Edward Cutrell.
\newblock Enable blind users’ experience in 3d virtual environments: The
  scene weaver prototype.
\newblock In \emph{Extended Abstracts of the 2023 CHI Conference on Human
  Factors in Computing Systems}, CHI EA '23, New York, NY, USA, 2023.
  Association for Computing Machinery.
\newblock ISBN 9781450394222.
\newblock \doi{10.1145/3544549.3583909}.
\newblock URL \url{https://doi.org/10.1145/3544549.3583909}.

\bibitem[Bigham et~al.(2010)Bigham, Jayant, Ji, Little, Miller, Miller, Miller,
  Tatarowicz, White, White, and Yeh]{bigham-vizwiz}
Jeffrey~P. Bigham, Chandrika Jayant, Hanjie Ji, Greg Little, Andrew Miller,
  Robert~C. Miller, Robin Miller, Aubrey Tatarowicz, Brandyn White, Samual
  White, and Tom Yeh.
\newblock Vizwiz: Nearly real-time answers to visual questions.
\newblock In \emph{Proceedings of the 23nd Annual ACM Symposium on User
  Interface Software and Technology}, UIST '10, page 333–342, New York, NY,
  USA, 2010. Association for Computing Machinery.
\newblock ISBN 9781450302715.
\newblock \doi{10.1145/1866029.1866080}.
\newblock URL \url{https://doi.org/10.1145/1866029.1866080}.

\bibitem[Brady et~al.(2013)Brady, Morris, Zhong, White, and
  Bigham]{brady-visual-challenges-everyday-lives}
Erin Brady, Meredith~Ringel Morris, Yu~Zhong, Samuel White, and Jeffrey~P.
  Bigham.
\newblock Visual challenges in the everyday lives of blind people.
\newblock In \emph{Proceedings of the SIGCHI Conference on Human Factors in
  Computing Systems}, CHI '13, page 2117–2126, New York, NY, USA, 2013.
  Association for Computing Machinery.
\newblock ISBN 9781450318990.
\newblock \doi{10.1145/2470654.2481291}.
\newblock URL \url{https://doi.org/10.1145/2470654.2481291}.

\bibitem[Braun and Clarke(2008)]{braun_using_2008}
Virginia Braun and Victoria Clarke.
\newblock Using thematic analysis in psychology.
\newblock \emph{Qualitative Research in Psychology}, July 2008.
\newblock \doi{10.1191/1478088706qp063oa}.
\newblock URL
  \url{https://www.tandfonline.com/doi/abs/10.1191/1478088706qp063oa}.
\newblock Publisher: Taylor \& Francis Group.

\bibitem[Charts()]{steam-charts-top-played-games}
Steam Charts.
\newblock Steamcharts - tracking what's played.
\newblock URL \url{https://steamcharts.com/top}.

\bibitem[De~Pascale et~al.(2008)De~Pascale, Mulatto, and
  Prattichizzo]{depascale-haptics-to-secondlife}
Maurizio De~Pascale, Sara Mulatto, and Domenico Prattichizzo.
\newblock Bringing haptics to second life for visually impaired people.
\newblock In \emph{Haptics: Perception, Devices and Scenarios: 6th
  International Conference, EuroHaptics 2008 Madrid, Spain, June 10-13, 2008
  Proceedings 6}, pages 896--905. Springer, 2008.
\newblock \doi{10.1007/978-3-540-69057-3_112}.
\newblock URL \url{https://doi.org/10.1007/978-3-540-69057-3_112}.

\bibitem[Eyes({\natexlab{a}})]{be-my-eyes-homepage}
Be~My Eyes.
\newblock Be my eyes - see the world together, {\natexlab{a}}.
\newblock URL \url{https://www.bemyeyes.com}.

\bibitem[Eyes({\natexlab{b}})]{be-my-eyes-sighted-volunteer}
Be~My Eyes.
\newblock Sighted volunteer - be my eyes help center, {\natexlab{b}}.
\newblock URL
  \url{https://support.bemyeyes.com/hc/en-us/categories/360000920938-Sighted-Volunteer}.

\bibitem[Fingas()]{fingas-how-does-alexa-work}
Roger Fingas.
\newblock How does alexa work? the tech behind amazon's virtual assistant,
  explained.
\newblock URL
  \url{https://www.androidauthority.com/how-does-alexa-work-3209316/}.

\bibitem[Gluck and Brinkley(2020)]{gluck-the-enclosing-dark}
Aaron Gluck and Julian Brinkley.
\newblock Implementing 'the enclosing dark': A vr auditory adventure.
\newblock \emph{Journal on Technology and Persons with Disabilities},
  8:\penalty0 149--159, 2020.
\newblock URL \url{http://hdl.handle.net/10211.3/215985}.

\bibitem[Gluck et~al.(2021)Gluck, Boateng, and
  Brinkley]{gluck-racing-in-the-dark}
Aaron Gluck, Kwajo Boateng, and Julian Brinkley.
\newblock Racing in the dark: Exploring accessible virtual reality by
  developing a racing game for people who are blind.
\newblock \emph{Proceedings of the Human Factors and Ergonomics Society Annual
  Meeting}, 65\penalty0 (1):\penalty0 1114--1118, 2021.
\newblock \doi{10.1177/1071181321651224}.
\newblock URL \url{https://doi.org/10.1177/1071181321651224}.

\bibitem[Gon{\c{c}}alves et~al.(2023{\natexlab{a}})Gon{\c{c}}alves,
  Pi{\c{c}}arra, Pais, Guerreiro, and Rodrigues]{gonccalves2023my}
David Gon{\c{c}}alves, Manuel Pi{\c{c}}arra, Pedro Pais, Jo{\~a}o Guerreiro,
  and Andr{\'e} Rodrigues.
\newblock " my zelda cane": Strategies used by blind players to play
  visual-centric digital games.
\newblock In \emph{Proceedings of the 2023 CHI conference on human factors in
  computing systems}, pages 1--15, 2023{\natexlab{a}}.

\bibitem[Gon{\c{c}}alves et~al.(2023{\natexlab{b}})Gon{\c{c}}alves, Rodrigues,
  Guerreiro, and Guerreiro]{gonccalves2023inclusive}
In{\^e}s Gon{\c{c}}alves, Andr{\'e} Rodrigues, Tiago Guerreiro, and Jo{\~a}o
  Guerreiro.
\newblock Inclusive social virtual environments: Exploring the acceptability of
  different navigation and awareness techniques.
\newblock In \emph{Extended Abstracts of the 2023 CHI Conference on Human
  Factors in Computing Systems}, pages 1--7, 2023{\natexlab{b}}.
\newblock \doi{10.1145/3544549.3585700}.
\newblock URL \url{https://doi.org/10.1145/3544549.3585700}.

\bibitem[Gonzalez~Penuela et~al.(2022)Gonzalez~Penuela, Poremba, Trice, and
  Azenkot]{penuela-hands-on}
Ricardo~E. Gonzalez~Penuela, Wren Poremba, Christina Trice, and Shiri Azenkot.
\newblock Hands-on: Using gestures to control descriptions of a virtual
  environment for people with visual impairments.
\newblock In \emph{Adjunct Proceedings of the 35th Annual ACM Symposium on User
  Interface Software and Technology}, UIST '22 Adjunct, New York, NY, USA,
  2022. Association for Computing Machinery.
\newblock ISBN 9781450393218.
\newblock \doi{10.1145/3526114.3558669}.
\newblock URL \url{https://doi.org/10.1145/3526114.3558669}.

\bibitem[Greener()]{greener-second-life-traffic}
Rory Greener.
\newblock Second life storefront user traffic jumps 35 percent in 2021.
\newblock URL
  \url{https://www.xrtoday.com/virtual-reality/second-life-user-traffic-jumps-35-percent-in-2021/}.

\bibitem[Guerreiro et~al.(2023)Guerreiro, Kim, Nogueira, Chung, Rodrigues, and
  Oh]{guerreiro2023design}
Jo{\~a}o Guerreiro, Yujin Kim, Rodrigo Nogueira, SeungA Chung, Andr{\'e}
  Rodrigues, and Uran Oh.
\newblock The design space of the auditory representation of objects and their
  behaviours in virtual reality for blind people.
\newblock \emph{IEEE Transactions on Visualization and Computer Graphics},
  29\penalty0 (5):\penalty0 2763--2773, 2023.

\bibitem[Guntalilib()]{guntalilib-screenreaders-vr}
Rhea~Althea Guntalilib.
\newblock Screenreader experience of a virtual reality conference.
\newblock URL
  \url{https://equalentry.com/screenreader-review-of-virtual-reality-conference-technology/}.

\bibitem[Hao et~al.(2023)Hao, Feng, Rizzo, Wang, and
  Fang]{hao-detect-and-approach-navigation}
Yu~Hao, Junchi Feng, John-Ross Rizzo, Yao Wang, and Yi~Fang.
\newblock Detect and approach: Close-range navigation support for people with
  blindness and low vision.
\newblock In \emph{European Conference on Computer Vision}, pages 607--622.
  Springer, 2023.
\newblock URL \url{https://doi.org/10.48550/arXiv.2208.08477}.

\bibitem[Huang(2009)]{huang-3d-haptic-labyrinth}
Yingying Huang.
\newblock Exploration on interface usability in a haptic 3d virtual labyrinth
  for visually impaired users.
\newblock In \emph{IADIS International Conference Interfaces and Human Computer
  Interaction}, 2009.
\newblock URL
  \url{https://www.csc.kth.se/~yngve/YingyingThesis/D-IHCI2009-publ.pdf}.

\bibitem[Institute()]{braille-institute-human-guide-techniques}
Braille Institute.
\newblock Braille institute human guide techniques.
\newblock URL
  \url{https://www.brailleinstitute.org/wp-content/uploads/2020/06/Braille-Institute-Human-Guide-Techniques-Accessible-Final-2020-0612.pdf}.

\bibitem[Ji et~al.(2022)Ji, Cochran, and Zhao]{ji-vrbubble}
Tiger~F. Ji, Brianna Cochran, and Yuhang Zhao.
\newblock Vrbubble: Enhancing peripheral awareness of avatars for people with
  visual impairments in social virtual reality.
\newblock In \emph{The 24th International {ACM} {SIGACCESS} Conference on
  Computers and Accessibility}. ACM, oct 2022.
\newblock \doi{10.1145/3517428.3544821}.
\newblock URL \url{https://doi.org/10.11452F3517428.3544821}.

\bibitem[Kaltenhauser and
  Schöning(2023)]{kaltenhauser-ghosts-from-past-secondlife}
Annika Kaltenhauser and Johannes Schöning.
\newblock Reawakening the ghosts from the past? accessibility lessons learned
  from second life.
\newblock 04 2023.
\newblock URL \url{https://www.alexandria.unisg.ch/269327/}.

\bibitem[Khalifa and Roxley()]{khalifa-radegast}
Latif Khalifa and Cinder Roxley.
\newblock Radegast github repository.
\newblock URL \url{https://github.com/cinderblocks/radegast}.

\bibitem[Kim(2020)]{kim-vivr-cane}
Jinmo Kim.
\newblock Vivr: Presence of immersive interaction for visual impairment virtual
  reality.
\newblock \emph{IEEE Access}, 8:\penalty0 196151--196159, 2020.
\newblock \doi{10.1109/ACCESS.2020.3034363}.
\newblock URL \url{https://doi.org/10.1109/ACCESS.2020.3034363}.

\bibitem[Kornbrot et~al.(2007)Kornbrot, Penn, Petrie, Furner, and
  Hardwick]{kornbrot-roughness}
Diana Kornbrot, Paul Penn, Helen Petrie, Stephen Furner, and Andrew Hardwick.
\newblock Roughness perception in haptic virtual reality for sighted and blind
  people.
\newblock \emph{Perception \& psychophysics}, 69:\penalty0 502--512, 2007.
\newblock URL \url{https://doi.org/10.3758/BF03193907}.

\bibitem[Lakhani et~al.(2022)Lakhani, Lakhotiya, and
  Mulla]{neha-be-my-eyes-aid-for-vi}
Neha Lakhani, Harshita Lakhotiya, and Nikahat Mulla.
\newblock Be my eyes: An aid for the visually impaired.
\newblock In \emph{2022 IEEE 3rd Global Conference for Advancement in
  Technology (GCAT)}, pages 1--6, 2022.
\newblock \doi{10.1109/GCAT55367.2022.9972160}.
\newblock URL \url{https://doi.org/10.1109/GCAT55367.2022.9972160}.

\bibitem[Lee et~al.(2018)Lee, Reddie, Gurdasani, Wang, Beck, Rosson, and
  Carroll]{lee-conversations-for-vision}
Sooyeon Lee, Madison Reddie, Krish Gurdasani, Xiying Wang, Jordan Beck,
  Mary~Beth Rosson, and John~M. Carroll.
\newblock Conversations for vision: Remote sighted assistants helping people
  with visual impairments.
\newblock \emph{CoRR}, abs/1812.00148, 2018.
\newblock \doi{10.48550/arXiv.1812.00148}.
\newblock URL \url{http://arxiv.org/abs/1812.00148}.

\bibitem[Lee et~al.(2020)Lee, Reddie, Tsai, Beck, Rosson, and
  Carroll]{lee-emerging-remote-sighted-assistance}
Sooyeon Lee, Madison Reddie, Chun-Hua Tsai, Jordan Beck, Mary~Beth Rosson, and
  John~M. Carroll.
\newblock The emerging professional practice of remote sighted assistance for
  people with visual impairments.
\newblock In \emph{Proceedings of the 2020 CHI Conference on Human Factors in
  Computing Systems}, CHI '20, page 1–12, New York, NY, USA, 2020.
  Association for Computing Machinery.
\newblock ISBN 9781450367080.
\newblock \doi{10.1145/3313831.3376591}.
\newblock URL \url{https://doi.org/10.1145/3313831.3376591}.

\bibitem[Li et~al.(2023)Li, Shinohara, and Peiris]{li-soundvizvr}
Ziming Li, Kristen Shinohara, and Roshan~L Peiris.
\newblock Exploring the use of the soundvizvr plugin with game developers in
  the development of sound-accessible virtual reality games.
\newblock CHI EA '23, New York, NY, USA, 2023. Association for Computing
  Machinery.
\newblock ISBN 9781450394222.
\newblock \doi{10.1145/3544549.3585750}.
\newblock URL \url{https://doi.org/10.1145/3544549.3585750}.

\bibitem[Lumbreras and Sánchez(1999)]{audiodoom}
Maruricio Lumbreras and Jaime Sánchez.
\newblock Interactive {3D} sound hyperstories for blind children.
\newblock In \emph{Proceedings of the {SIGCHI} conference on {Human} factors in
  computing systems the {CHI} is the limit - {CHI} '99}, pages 318--325,
  Pittsburgh, Pennsylvania, United States, 1999. ACM Press.
\newblock ISBN 978-0-201-48559-2.
\newblock \doi{10.1145/302979.303101}.
\newblock URL \url{https://dl.acm.org/doi/10.1145/302979.303101}.

\bibitem[McVeigh-Schultz et~al.(2018)McVeigh-Schultz, M\'{a}rquez~Segura,
  Merrill, and Isbister]{mcveigh-schultz-mapping-social-vr-design-ecology}
Joshua McVeigh-Schultz, Elena M\'{a}rquez~Segura, Nick Merrill, and Katherine
  Isbister.
\newblock What's it mean to "be social" in vr? mapping the social vr design
  ecology.
\newblock In \emph{Proceedings of the 2018 ACM Conference Companion Publication
  on Designing Interactive Systems}, DIS '18 Companion, page 289–294, New
  York, NY, USA, 2018. Association for Computing Machinery.
\newblock ISBN 9781450356312.
\newblock \doi{10.1145/3197391.3205451}.
\newblock URL \url{https://doi.org/10.1145/3197391.3205451}.

\bibitem[Meta({\natexlab{a}})]{meta-haptics-studio}
Meta.
\newblock Haptics studio, {\natexlab{a}}.
\newblock URL
  \url{https://developer.oculus.com/experimental/exp-haptics-studio/}.

\bibitem[Meta({\natexlab{b}})]{meta-health-and-safety}
Meta.
\newblock Meta quest health \& safety - learn how to stay safe while using your
  meta quest, {\natexlab{b}}.
\newblock URL \url{https://www.oculus.com/safety-center/quest/}.

\bibitem[Metaverse and Care?()]{what-is-the-metaverse}
What Is~The Metaverse and Why Should~You Care?
\newblock What is the metaverse and why should you care?
\newblock URL
  \url{https://www.forbes.com/sites/deborahlovich/2022/05/11/what-is-the-metaverse-and-why-should-you-care/}.

\bibitem[Morelli et~al.(2010)Morelli, Foley, Columna, Lieberman, and
  Folmer]{morelli_vi-tennis_2010}
Tony Morelli, John Foley, Luis Columna, Lauren Lieberman, and Eelke Folmer.
\newblock {VI}-{Tennis}: a vibrotactile/audio exergame for players who are
  visually impaired.
\newblock In \emph{Proceedings of the {Fifth} {International} {Conference} on
  the {Foundations} of {Digital} {Games}}, {FDG} '10, pages 147--154, New York,
  NY, USA, June 2010. Association for Computing Machinery.
\newblock ISBN 978-1-60558-937-4.
\newblock \doi{10.1145/1822348.1822368}.
\newblock URL \url{https://doi.org/10.1145/1822348.1822368}.

\bibitem[Nair et~al.(2021)Nair, Karp, Silverman, Kalra, Lehv, Jamil, and
  Smith]{nair-navstick}
Vishnu Nair, Jay~L. Karp, Samuel Silverman, Mohar Kalra, Hollis Lehv, Faizan
  Jamil, and Brian~A. Smith.
\newblock Navstick: Making video games blind-accessible via the ability to look
  around.
\newblock \emph{CoRR}, abs/2109.01202, 2021.
\newblock \doi{10.1145/3472749.3474768}.
\newblock URL \url{https://arxiv.org/abs/2109.01202}.

\bibitem[Nair et~al.(2022)Nair, Ma, Gonzalez~Penuela, He, Lin, Hayes,
  Huddleston, Donnelly, and Smith]{nair2022uncovering}
Vishnu Nair, Shao-en Ma, Ricardo~E Gonzalez~Penuela, Yicheng He, Karen Lin,
  Mason Hayes, Hannah Huddleston, Matthew Donnelly, and Brian~A Smith.
\newblock Uncovering visually impaired gamers’ preferences for spatial
  awareness tools within video games.
\newblock In \emph{Proceedings of the 24th International ACM SIGACCESS
  Conference on Computers and Accessibility}, pages 1--16, 2022.

\bibitem[Nguyen et~al.(2018)Nguyen, Kim, Park, Chen, Chen, Van~Fossan, and
  Chao]{nguyen-improved-quality-of-life-aira}
Brian~J. Nguyen, Yeji Kim, Kathryn Park, Allison~J. Chen, Scarlett Chen, Donald
  Van~Fossan, and Daniel~L. Chao.
\newblock {Improvement in Patient-Reported Quality of Life Outcomes in Severely
  Visually Impaired Individuals Using the Aira Assistive Technology System}.
\newblock \emph{Translational Vision Science \& Technology}, 7\penalty0
  (5):\penalty0 30--30, 10 2018.
\newblock ISSN 2164-2591.
\newblock \doi{10.1167/tvst.7.5.30}.
\newblock URL \url{https://doi.org/10.1167/tvst.7.5.30}.

\bibitem[Nikolakis et~al.(2004)Nikolakis, Tzovaras, Moustakidis, and
  Strintzis]{nikolakis-cybergrasp}
Georgios Nikolakis, Dimitrios Tzovaras, Serafim Moustakidis, and Michael~G
  Strintzis.
\newblock Cybergrasp and phantom integration: Enhanced haptic access for
  visually impaired users.
\newblock In \emph{9th Conference Speech and Computer}, 2004.
\newblock URL
  \url{https://www.isca-speech.org/archive_open/specom_04/spc4_507.pdf}.

\bibitem[of~the Blind and
  Impaired()]{wisconsin-council-household-products-labeling}
Wisconsin~Council of~the Blind and Visually Impaired.
\newblock Identifying household products through labeling.
\newblock URL
  \url{https://wcblind.org/2019/11/identifying-household-products-through-labeling/}.

\bibitem[Oktay and folmer(2010)]{oktay-textsl-secondlife}
Bugra Oktay and eelke folmer.
\newblock Textsl: A screen reader accessible interface for second life.
\newblock page~21, 04 2010.
\newblock \doi{10.1145/1805986.1806017}.
\newblock URL \url{https://doi.org/10.1145/1805986.1806017}.

\bibitem[Populations()]{mmo-population-list}
MMO Populations.
\newblock Top mmos webpage, most popular mmos server population \& player
  count.
\newblock URL \url{https://mmo-population.com/list}.

\bibitem[Post()]{ar-post-best-vr-for-socializing}
AR~Post.
\newblock The best vr apps for socializing with friends.
\newblock URL
  \url{https://arpost.co/2022/07/29/best-vr-apps-for-socializing-with-friends/}.

\bibitem[Power et~al.(2012)Power, Freire, Petrie, and
  Swallow]{power2012guidelines}
Christopher Power, Andr{\'e} Freire, Helen Petrie, and David Swallow.
\newblock Guidelines are only half of the story: accessibility problems
  encountered by blind users on the web.
\newblock In \emph{Proceedings of the SIGCHI conference on human factors in
  computing systems}, pages 433--442, 2012.

\bibitem[Salisbury et~al.(2017)Salisbury, Kamar, and
  Morris]{salisbury-conversational-crowdsourcing}
Elliot Salisbury, Ece Kamar, and Meredith Morris.
\newblock Toward scalable social alt text: Conversational crowdsourcing as a
  tool for refining vision-to-language technology for the blind.
\newblock \emph{Proceedings of the AAAI Conference on Human Computation and
  Crowdsourcing}, 5\penalty0 (1):\penalty0 147--156, Sep. 2017.
\newblock \doi{10.1609/hcomp.v5i1.13301}.
\newblock URL \url{https://ojs.aaai.org/index.php/HCOMP/article/view/13301}.

\bibitem[Services()]{helen-keller-human-guide}
Helen Keller National~Center Services.
\newblock Human guide techniques.
\newblock URL
  \url{https://www.helenkeller.org/resources/human-guide-techniques/}.

\bibitem[Sight()]{boston-sight-safe-home}
Boston Sight.
\newblock How to create a safe home for the visually impaired and totally
  blind.
\newblock URL
  \url{https://www.bostonsight.org/how-to-create-a-safe-home-for-the-visually-impaired-and-totally-blind/}.

\bibitem[Sinclair et~al.(2019)Sinclair, Ofek, Gonzalez-Franco, and
  Holz]{sinclair-capstan-crunch}
Mike Sinclair, Eyal Ofek, Mar Gonzalez-Franco, and Christian Holz.
\newblock Capstancrunch: A haptic vr controller with user-supplied force
  feedback.
\newblock In \emph{Proceedings of the 32nd Annual ACM Symposium on User
  Interface Software and Technology}, UIST '19, page 815–829, New York, NY,
  USA, 2019. Association for Computing Machinery.
\newblock ISBN 9781450368162.
\newblock \doi{10.1145/3332165.3347891}.
\newblock URL \url{https://doi.org/10.1145/3332165.3347891}.

\bibitem[Siu et~al.(2020)Siu, Sinclair, Kovacs, Ofek, Holz, and
  Cutrell]{siu-VR-without-vision}
Alexa~F. Siu, Mike Sinclair, Robert Kovacs, Eyal Ofek, Christian Holz, and
  Edward Cutrell.
\newblock Virtual reality without vision: A haptic and auditory white cane to
  navigate complex virtual worlds.
\newblock In \emph{Proceedings of the 2020 CHI Conference on Human Factors in
  Computing Systems}, CHI '20, page 1–13, New York, NY, USA, 2020.
  Association for Computing Machinery.
\newblock ISBN 9781450367080.
\newblock \doi{10.1145/3313831.3376353}.
\newblock URL \url{https://doi.org/10.1145/3313831.3376353}.

\bibitem[Sun et~al.(2023)Sun, Stellmacher, Kaltenhauser, Wagener, Neumann, and
  Sch{\"o}ning]{sun2023alt}
Yu~Sun, Carolin Stellmacher, Annika Kaltenhauser, Nadine Wagener, Daniel
  Neumann, and Johannes Sch{\"o}ning.
\newblock Alt text and alt sense in vr: Engaging screen reader users within the
  metaverse through multisenses.
\newblock 2023.

\bibitem[Team()]{xr-today-best-social-vr}
XR~Today Team.
\newblock The best social apps in vr.
\newblock URL
  \url{https://www.xrtoday.com/virtual-reality/the-best-social-apps-in-vr/}.

\bibitem[Trewin et~al.(2008)Trewin, Hanson, Laff, and Cavender]{trewin-powerup}
Shari Trewin, Vicki~L. Hanson, Mark~R. Laff, and Anna Cavender.
\newblock Powerup: An accessible virtual world.
\newblock In \emph{Proceedings of the 10th International ACM SIGACCESS
  Conference on Computers and Accessibility}, Assets '08, page 177–184, New
  York, NY, USA, 2008. Association for Computing Machinery.
\newblock ISBN 9781595939760.
\newblock \doi{10.1145/1414471.1414504}.
\newblock URL \url{https://doi.org/10.1145/1414471.1414504}.

\bibitem[Tzovaras et~al.(2009)Tzovaras, Moustakas, Nikolakis, and
  Strintzis]{tzovaras-mr-white-cane}
Dimitrios Tzovaras, Konstantinos Moustakas, Georgios Nikolakis, and Michael~G
  Strintzis.
\newblock Interactive mixed reality white cane simulation for the training of
  the blind and the visually impaired.
\newblock \emph{Personal and Ubiquitous Computing}, 13:\penalty0 51--58, 2009.
\newblock \doi{10.1007/s00779-007-0171-2}.
\newblock URL \url{https://doi.org/10.1007/s00779-007-0171-2}.

\bibitem[Wedoff et~al.(2019)Wedoff, Ball, Wang, Khoo, Lieberman, and
  Rector]{wedoff-virtualshowdown}
Ryan Wedoff, Lindsay Ball, Amelia Wang, Yi~Xuan Khoo, Lauren Lieberman, and
  Kyle Rector.
\newblock Virtual showdown: An accessible virtual reality game with scaffolds
  for youth with visual impairments.
\newblock In \emph{Proceedings of the 2019 CHI Conference on Human Factors in
  Computing Systems}, CHI '19, page 1–15, New York, NY, USA, 2019.
  Association for Computing Machinery.
\newblock ISBN 9781450359702.
\newblock \doi{10.1145/3290605.3300371}.
\newblock URL \url{https://doi.org/10.1145/3290605.3300371}.

\bibitem[Westin(2004)]{westin-terraformers}
Thomas Westin.
\newblock Game accessibility case study : Terraformers - a real-time 3d graphic
  game.
\newblock 2004.
\newblock URL
  \url{https://www.researchgate.net/publication/250823995_Game_accessibility_case_study_Terraformers_-_a_real-time_3D_graphic_game}.

\bibitem[White et~al.(2008)White, Fitzpatrick, and
  McAllister]{white-towards-accessible-secondlife}
Gareth~R. White, Geraldine Fitzpatrick, and Graham McAllister.
\newblock Toward accessible 3d virtual environments for the blind and visually
  impaired.
\newblock In \emph{Proceedings of the 3rd International Conference on Digital
  Interactive Media in Entertainment and Arts}, DIMEA '08, page 134–141, New
  York, NY, USA, 2008. Association for Computing Machinery.
\newblock ISBN 9781605582481.
\newblock \doi{10.1145/1413634.1413663}.
\newblock URL \url{https://doi.org/10.1145/1413634.1413663}.

\bibitem[Wieland et~al.(2022)Wieland, Thevin, Schmidt, and
  Machulla]{wieland-nvc-joint-attention-guidelines}
Markus Wieland, Lauren Thevin, Albrecht Schmidt, and Tonja Machulla.
\newblock Non-verbal communication and joint attention between people with and
  without visual impairments: Deriving guidelines for inclusive conversations
  in virtual realities.
\newblock In \emph{Computers Helping People with Special Needs: 18th
  International Conference, ICCHP-AAATE 2022, Lecco, Italy, July 11–15, 2022,
  Proceedings, Part I}, page 295–304, Berlin, Heidelberg, 2022.
  Springer-Verlag.
\newblock ISBN 978-3-031-08647-2.
\newblock \doi{10.1007/978-3-031-08648-9_34}.
\newblock URL \url{https://doi.org/10.1007/978-3-031-08648-9_34}.

\bibitem[Winlock et~al.(2010)Winlock, Christiansen, and
  Belongie]{winlock-realtime-grocery-detection}
Tess Winlock, Eric Christiansen, and Serge Belongie.
\newblock Toward real-time grocery detection for the visually impaired.
\newblock 06 2010.
\newblock \doi{10.1109/CVPRW.2010.5543576}.
\newblock URL \url{https://doi.org/10.1109/CVPRW.2010.5543576}.

\bibitem[Zhang et~al.(2022)Zhang, Deldari, Lu, Yao, and
  Zhao]{zhang-part-of-me-avatar-diversity}
Kexin Zhang, Elmira Deldari, Zhicong Lu, Yaxing Yao, and Yuhang Zhao.
\newblock “it’s just part of me:” understanding avatar diversity and
  self-presentation of people with disabilities in social virtual reality.
\newblock In \emph{Proceedings of the 24th International ACM SIGACCESS
  Conference on Computers and Accessibility}, ASSETS '22, New York, NY, USA,
  2022. Association for Computing Machinery.
\newblock ISBN 9781450392587.
\newblock \doi{10.1145/3517428.3544829}.
\newblock URL \url{https://doi.org/10.1145/3517428.3544829}.

\bibitem[Zhang et~al.(2020)Zhang, Wu, Yang, Tang, and
  Zhu]{zhang-exploring-ve-mixed-reality-cane}
Lei Zhang, Klevin Wu, Bin Yang, Hao Tang, and Zhigang Zhu.
\newblock Exploring virtual environments by visually impaired using a mixed
  reality cane without visual feedback.
\newblock In \emph{2020 IEEE International Symposium on Mixed and Augmented
  Reality Adjunct (ISMAR-Adjunct)}, pages 51--56, 2020.
\newblock \doi{10.1109/ISMAR-Adjunct51615.2020.00028}.
\newblock URL \url{https://doi.org/10.1109/ISMAR-Adjunct51615.2020.00028}.

\bibitem[Zhao et~al.(2016)Zhao, Szpiro, Knighten, and Azenkot]{zhao-cue-see}
Yuhang Zhao, Sarit Szpiro, Jonathan Knighten, and Shiri Azenkot.
\newblock Cuesee: Exploring visual cues for people with low vision to
  facilitate a visual search task.
\newblock In \emph{Proceedings of the 2016 ACM International Joint Conference
  on Pervasive and Ubiquitous Computing}, UbiComp '16, page 73–84, New York,
  NY, USA, 2016. Association for Computing Machinery.
\newblock ISBN 9781450344616.
\newblock \doi{10.1145/2971648.2971730}.
\newblock URL \url{https://doi.org/10.1145/2971648.2971730}.

\bibitem[Zhao et~al.(2018)Zhao, Bennett, Benko, Cutrell, Holz, Morris, and
  Sinclair]{zhao-Canetroller}
Yuhang Zhao, Cynthia~L. Bennett, Hrvoje Benko, Edward Cutrell, Christian Holz,
  Meredith~Ringel Morris, and Mike Sinclair.
\newblock Enabling people with visual impairments to navigate virtual reality
  with a haptic and auditory cane simulation.
\newblock In \emph{Proceedings of the 2018 CHI Conference on Human Factors in
  Computing Systems}, CHI '18, page 1–14, New York, NY, USA, 2018.
  Association for Computing Machinery.
\newblock ISBN 9781450356206.
\newblock \doi{10.1145/3173574.3173690}.
\newblock URL \url{https://doi.org/10.1145/3173574.3173690}.

\bibitem[Zhao et~al.(2019)Zhao, Cutrell, Holz, Morris, Ofek, and
  Wilson]{zhao2019seeingvr}
Yuhang Zhao, Ed~Cutrell, Christian Holz, Meredith~Ringel Morris, Eyal Ofek, and
  Andy Wilson.
\newblock Seeingvr: A set of tools to make virtual reality more accessible to
  people with low vision.
\newblock In \emph{CHI 2019}. ACM, May 2019.
\newblock \doi{10.1145/3290605.3300341}.
\newblock URL
  \url{https://www.microsoft.com/en-us/research/publication/seeingvr-a-set-of-tools-to-make-virtual-reality-more-accessible-to-people-with-low-vision-2/}.

\end{thebibliography}

\appendix

\end{document}